\documentclass[reprint,prb,twocolumn,showpacs,superscriptaddress,aps]{revtex4-1}

\usepackage{graphicx}
\DeclareGraphicsExtensions{.png,.jpg,.eps}

\usepackage{float}
\usepackage{xcolor}
\usepackage{amsmath}
\usepackage{caption}
\usepackage{graphics}
\usepackage{float}

\begin{document}	

\title{Absence of Ferromagnetism in VSe$_2$ Caused by its Charge Density Wave Phase}

\author{Adolfo O. Fumega}
  \email{adolfo.otero.fumega@usc.es}
 \affiliation{Departamento de F\'{i}sica Aplicada,
  Universidade de Santiago de Compostela, E-15782 Campus Sur s/n,
  Santiago de Compostela, Spain}
\affiliation{Instituto de Investigaci\'{o}ns Tecnol\'{o}xicas,
  Universidade de Santiago de Compostela, E-15782 Campus Sur s/n,
  Santiago de Compostela, Spain} 
 \author{M. Gobbi}
\affiliation{Centro de F\'{i}sica de Materiales (CSIC –UPV/EHU), Paseo Manuel de Lardiz\'{a}bal, 5 – E-20018 Donostia – San Sebasti\'{a}n Gipzukoa – Spain} 
  \author{P. Dreher}
\affiliation{Centro de F\'{i}sica de Materiales (CSIC –UPV/EHU), Paseo Manuel de Lardiz\'{a}bal, 5 – E-20018 Donostia – San Sebasti\'{a}n Gipzukoa – Spain} 
   \author{W. Wan}
\affiliation{Centro de F\'{i}sica de Materiales (CSIC –UPV/EHU), Paseo Manuel de Lardiz\'{a}bal, 5 – E-20018 Donostia – San Sebasti\'{a}n Gipzukoa – Spain} 
  \author{C. Gonz\'{a}lez-Orellana}
\affiliation{Centro de F\'{i}sica de Materiales (CSIC –UPV/EHU), Paseo Manuel de Lardiz\'{a}bal, 5 – E-20018 Donostia – San Sebasti\'{a}n Gipzukoa – Spain} 
  \author{M. Pe\~{n}a-D\'{i}az}
\affiliation{Centro de F\'{i}sica de Materiales (CSIC –UPV/EHU), Paseo Manuel de Lardiz\'{a}bal, 5 – E-20018 Donostia – San Sebasti\'{a}n Gipzukoa – Spain} 
 \author{C. Rogero}
\affiliation{Centro de F\'{i}sica de Materiales (CSIC –UPV/EHU), Paseo Manuel de Lardiz\'{a}bal, 5 – E-20018 Donostia – San Sebasti\'{a}n Gipzukoa – Spain} 
 \author{J. Herrero-Mart\'{i}n}
\affiliation{ALBA Synchrotron Light Source, Cerdanyola del Vall\`es, 08290 Barcelona, Catalonia, Spain}
\author{P. Gargiani}
\affiliation{ALBA Synchrotron Light Source, Cerdanyola del Vall\`es, 08290 Barcelona, Catalonia, Spain}
\author{M. Ilyn}
\affiliation{Centro de F\'{i}sica de Materiales (CSIC –UPV/EHU), Paseo Manuel de Lardiz\'{a}bal, 5 – E-20018 Donostia – San Sebasti\'{a}n Gipzukoa – Spain} 
   \author{M. M. Ugeda}
\affiliation{Centro de F\'{i}sica de Materiales (CSIC –UPV/EHU), Paseo Manuel de Lardiz\'{a}bal, 5 – E-20018 Donostia – San Sebasti\'{a}n Gipzukoa – Spain} 
    \affiliation{Donostia International Physics Center, DIPC, 20018 Donostia-San Sebastian, Basque Country, Spain}
  \affiliation{IKERBASQUE, Basque Foundation for Science, 48013 Bilbao, Basque Country, Spain}
\author{Victor Pardo}
  \email{victor.pardo@usc.es}
\affiliation{Departamento de F\'{i}sica Aplicada,
  Universidade de Santiago de Compostela, E-15782 Campus Sur s/n,
  Santiago de Compostela, Spain}
\affiliation{Instituto de Investigaci\'{o}ns Tecnol\'{o}xicas,
  Universidade de Santiago de Compostela, E-15782 Campus Sur s/n,
  Santiago de Compostela, Spain}  
  \author{S.~Blanco-Canosa}
\email{sblanco@dipc.org}
\affiliation{Donostia International Physics Center, DIPC, 20018 Donostia-San Sebastian, Basque Country, Spain}
\affiliation{IKERBASQUE, Basque Foundation for Science, 48013 Bilbao, Basque Country, Spain}


\begin{abstract} 

How magnetism emerges in low-dimensional materials such as transition metal dichalcogenides at the monolayer limit is still an open question. Herein, we present a comprehensive study of the magnetic properties of single crystal and monolayer VSe$_{2}$, both experimentally and \emph{ab initio}. Magnetometry, X-ray magnetic circular dichrosim (XMCD) and \emph{ab initio} calculations demonstrate that the charge density wave in bulk stoichiometric VSe$_{2.0}$ causes a structural distortion with a strong reduction in the density of sates at the Fermi level, prompting the system towards a non-magnetic state but on the verge of a ferromagnetic instability. In the monolayer limit, the structural rearrangement induces a Peierls distortion with the opening of an energy gap at the Fermi level and the absence of magnetic order. Control experiments on defect-induced VSe$_{2-\delta}$ single crystals show a breakdown of magnetism, discarding vacancies as a possible origin of magnetic order in VSe$_{2}$.

\end{abstract}

\maketitle

\section{\label{intro}INTRODUCTION}
Since the discovery of graphene \cite{Graphene}, much scientific effort is concentrated on the characterization of purely two-dimensional (2D) materials. In particular, the family of layered transition metal dichalcogenides (TMDs, MX$_2$: M=Nb, Ti, V,... X= S, Se, Te) is attracting great attention \cite{manzeli_2d_2017,CHOI2017116,C7TC01088E,XIA20171} since emergent phenomena driven by novel electronic, optical and quantum many-body properties at the 2D limit could lead to new applications \cite{Aki14}. Control over the material thickness down to the monolayer limit has been accurately achieved by mechanical exfoliation \cite{xu_ultrathin_2013}, chemical vapor deposition and layer-by-layer Molecular Beam Epitaxy (MBE) \cite{nano_monolayer,Umemoto2018}, revealing that collective quantum ground states, coherent modulation of electronic periodicities \cite{PRL_monoVSe2}, superconductivity \cite{Uge16}, optoelectronic and valley excitonic physics \cite{Tra19,Sey19,Jin19,Ale19} survive down to the atomic scale. 

Nevertheless, long-range magnetic ordering in low dimensions has been elusive for decades. Theoretically, the Mermin-Wagner theorem \cite{Mer66} prohibits long-range magnetic order in the 2D isotropic Heisenberg model at finite temperatures if the system is spin-rotational invariant. Nevertheless, Ising-type ferromagnetism has been observed in a purely 2D material \cite{huang_layer-dependent_2017,Bur18,Gib19}, paving the path for future spintronic applications. Moreover, simple defects in a host lattice strongly alter both macroscopic and local properties of the system. Magnetic order from disorder has been observed to emerge in superconducting cuprates upon substitution of non-magnetic ions by spinless impurities \cite{All09}, Kondo systems \cite{Pra17} and in highly oriented pyrolytic graphite (HOPG) \cite{Cer09}, demonstrating that grain boundaries, vacancies and point defects can act as magnetic nuclei in a non-magnetic matrix \cite{Yaz07}.

\begin{figure*}
  \centering
  \includegraphics[width=0.8\textwidth]{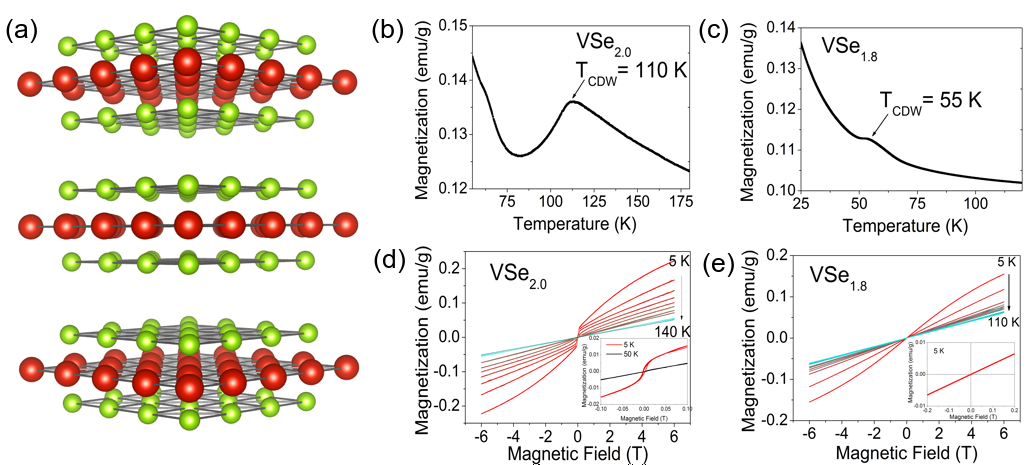}
     \caption{(a) Layered structure of VSe$_{2}$ in the \textit{P$\overline{3}$m1} space group. V atoms are represented as red spheres and Se atoms are shown in green. (b-c)  Temperature dependence of the magnetization for stoichiometric VSe$_{2.0}$ and Se-deficient VSe$_{1.8}$ single crystals. The charge density wave transition temperature (T$_{CDW}$) is signaled by a kink in the magnetization curve. (d-e) Magnetization vs field for VSe$_{2.0}$ and VSe$_{1.8}$ single crystals. Inset, zoom-in of the low field magnetization highlighting the small magnetic signal for VSe$_{2.0}$, not observed in VSe$_{1.8}$.}\label{magnetization}
\end{figure*}

Recently, a strong ferromagnetic  (FM)  signal at room temperature has been reported at the monolayer limit of the 2D van der Waals system VSe$_2$ \cite{bonilla_strong_2018}, broadening the sprectrum of 2D materials hosting magnetism in the ultrathin limit \cite{gong_discovery_2017,VS2_FM}. van der Waals stacked layers of VSe$_2$ consist of 6-fold coordinated V atoms crystallizing in a trigonal (1T) structure (space group \textit{P$\overline{3}$m1}  in the normal state, NS, see Fig. 1(a)).  As many TMDs \cite{TiSe2_CDW, TMD_CDW}, both transport \cite{van_bruggen_magnetic_1976,yadav_electronic_2010} and diffraction data \cite{williams_charge_1976} show that bulk VSe$_{2}$ develops a 3D charge density wave (CDW) below T$_{CDW}\sim110$ K \cite{VSe2_CDW1} with a new commensurate 4$a$ $\times$ 4$a$ $\times$ 3$c$ lattice periodicity \cite{van_bruggen_magnetic_1976, 0022-3719-10-14-013,PhysRevLett.109.086401}.  A pseudogap appears at the Fermi surface \cite{BAYARD1976325} and the system remains paramagnetic \cite{van_bruggen_magnetic_1976, BAYARD1976325,barua_signatures_2017,cao_defect_2017}. Remarkably, T$_{CDW}$ increases at the monolayer limit \cite{xu_ultrathin_2013} and a gap opens at the Fermi level \cite{Umemoto2018} with a controversial magnetic behavior for exfoliated \cite{xu_ultrathin_2013} and MBE-grown monolayers \cite{bonilla_strong_2018}. However, a consistent picture about the microscopic origin of magnetism in VSe$_{2}$ and the development of FM order in this non-magnetic material in 2D is still lacking. In part, this is a consequence of the proximity to electronic and magnetic instabilities which can balance the competition between ground states in the 2D limit. Angle resolved photoemission (ARPES) \cite{PRL_monoVSe2} and scanning tunneling microscopy (STM) \cite{nano_monolayer} have revealed an electronic reconstruction of single layer VSe$_{2}$ compared with the bulk counterpart, without a detectable FM exchange splitting, casting doubts on whether magnetism originates from an induced band structure spin splitting caused by dimensionality reduction or extrinsic defects come into play. Indeed, previous density functional theory (DFT) calculations \cite{li_versatile_2014, ma_evidence_2012, PhysRevB_DI, NJP} found a FM ground state for both bulk and  single-layer VSe$_{2}$. However, these DFT calculations do not take into account the effect of the CDW structure and hence, the electronic reconstruction and the effect of impurities are largely overlooked.

In order to shed light about the nature of the magnetic ground state in VSe$_{2}$, we have carried out a comprehensive theoretical and experimental study of bulk single crystal and monolayer VSe$_{2}$. We report that, due to the reduction of the density of states (DOS) at the Fermi level caused by the opening of the CDW pseudo-gap, bulk VSe$_{2}$ is close to a ferromagnetic instability, which cannot be induced by defects or vacancies. Besides, monolayer VSe$_{2}$ shows a Peierls-like distortion that opens a gap at the ultrathin limit, preventing the system to develop a long-range FM order, in agreement with recent reports \cite{PRL_monoVSe2}.

\section{\label{methods}EXPERIMENTAL AND COMPUTATIONAL METHODS} 

Single crystals of VSe$_{2.0}$ and VSe$_{1.8}$ were grown by chemical vapor deposition following previous reports \cite{yadav_electronic_2010}. 5\% excess of Se for VSe$_{2.0}$ and stoichiometric V:Se (1:2) for VSe$_{1.8}$ was used during the synthesis. V:Se ratio was measured by Energy-dispersive X-ray spectroscopy (EDX). Single layer VSe$_{2.0}$ was grown by MBE on epitaxial bilayer graphene on Silicon carbide (Si-C) and highly oriented pyrolytic graphite (HOPG) substrates in an ultrahigh vacuum chamber with a base pressure of $\sim$ 1$\times$10$^{-9}$ and 250$^{\circ}$C. After the growth, three minutes post-annealing in Se-rich atmosphere was carried out to fill in the Se vacancies. A Se capping layer was deposited after cooling to prevent oxidation and was \textit{in situ} evaporated for XMCD experiments. Magnetic measurements on Se-capped VSe$_2$ monolayers grown on diamagnetic bilayer graphene/SiC were carried out in a SQUID magnetometer up
to 7 Tesla. X-ray magnetic circular dichroism (XMCD) at the V \textit{L$_{2,3}$}-edge up to 6 T was performed at the BOREAS beamline at ALBA synchrotron \cite{Bar16}. Normal and grazing incidence geometries are referred to to 90º and 20º angle between the beam and sample surface, the latter being more sensitive to in-plane magnetization, and the magnetic field is always parallel to the beam direction. 
Cluster calculations were carried out using crystal field theory implemented in the QUANTY code \cite{Hav12,Lu14} for the atomic-like 2p${^6}$ - 3d${^\delta}$ $\to$ 2p${^5}$-3d${^{\delta+1}}$ transitions. 
First principles DFT \emph{ab initio} electronic structure calculations \cite{HK,KS} were performed using an all-electron full potential code ({\sc wien2k} \cite{WIEN2k}).
The exchange-correlation term for the bulk structure was the generalized gradient approximation (GGA) in the Perdew-Burke-Ernzerhof \cite{PBE} scheme. 
The LDA+U method was used for the 2D case \cite{LDAU}. The calculations were carried out with a converged \textit{k}-mesh and a value of \textit{R}$_{mt}$\textit{K}$_{max}$= 7.0 and a \textit{R}$_{mt}$ value of 2.12 a. u. for both V and Se. Transport properties were obtained with the \textit{BoltzTrap2} code \cite{Boltztrap2} using a denser \textit{k}-mesh, solving the Boltzmann transport equation within the constant scattering time approximation.

\section{\label{results}RESULTS}

The paper is organized as follows: first, we present the experimental and theoretical results obtained for bulk VSe$_{2}$, followed by showing the results for the system at the monolayer limit. 

\subsection{\label{bulk}Bulk VSe$_{2}$}

Plotted in Fig. \ref{magnetization} (b-c) are the temperature dependence of the magnetization for VSe$_{2.0}$ and VSe$_{1.8}$ single crystals. Following previous reports \cite{van_bruggen_magnetic_1976,yadav_electronic_2010}, the CDW transition is identified as a kink in the magnetic susceptibility; at 110 K for VSe$_{2.0}$ and 55 K for VSe$_{1.8}$, evidencing the drop in transition temperature upon introducing Se defects. Interestingly, the field dependence of the magnetization shows s-shape magnetic behavior for VSe$_{2.0}$ (Fig. \ref{magnetization}d and inset) but absent in single crystals of VSe$_{1.8}$ (Fig. \ref{magnetization}e). Nevertheless, neither the magnetization of VSe$_{2.0}$ nor VSe$_{1.8}$ saturates at 6 T. The small magnetic behavior observed in VSe$_{2.0}$ could arise from Kondo impurities \cite{Bar17} or phase slippage of the CDW \cite{Gru94}.

\begin{figure}[!h]
  \centering
  \includegraphics[width=0.8\columnwidth]%
    {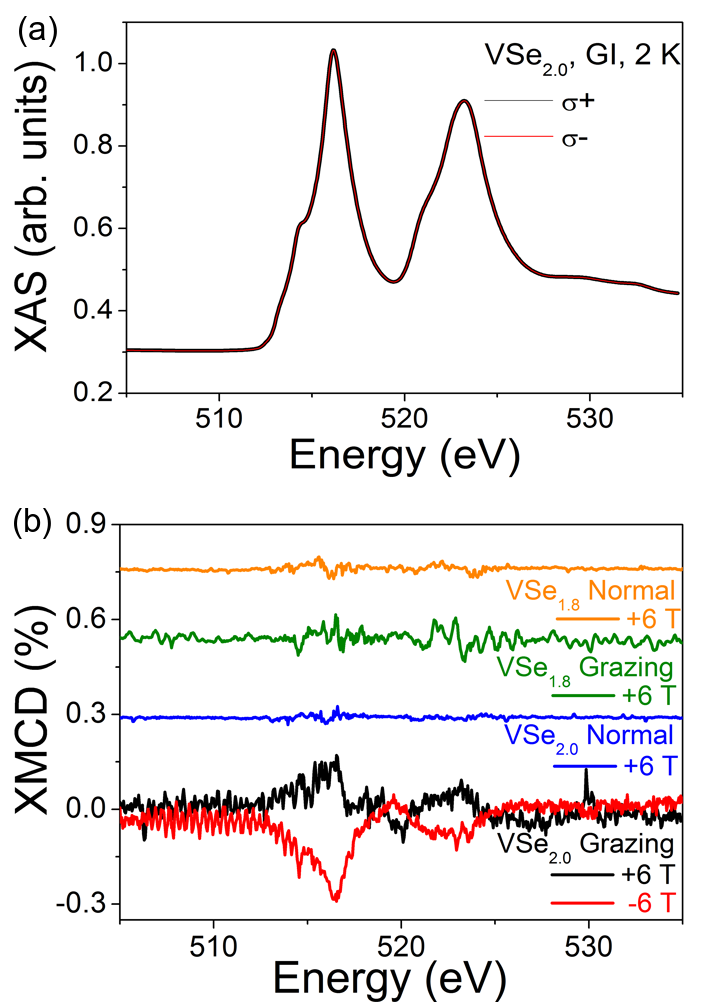}
     \caption{(a) XAS spectra for single crystal VSe$_{2.0}$ with circular positive $\sigma$$^+$ and $\sigma$$^-$ polarized light at 2 K, 6 T and grazing incidence geometry (70$^{\circ}$ off with respect to the c-axis). (b) XMCD spectra for VSe$_{2.0}$ and VSe$_{1.8}$ single crystals at 2 K, 6 T and normal (NI) and grazing (GI) incidence geometries. Only a small dichroic signal is discernible at GI.}\label{xas}
\end{figure}

To identify the source of magnetism in VSe$_{2.0}$ single crystals, we have carried out X-ray magnetic circular dichroism (XMCD) at the V \textit{L$_{2,3}$} edge at 6 T (Fig. \ref{xas}). As shown in Fig. \ref{xas}(b), a small XMCD signal, defined as ($\sigma$${^+}$-$\sigma$${^-}$), can only be detected in VSe$_{2.0}$ at the \textit{L$_{2,3}$} in grazing incidence geometry (GI) at 6 T, suggesting that small moments are in-plane aligned. On the other hand, no significant dichroic signal is detected in VSe$_{1.8}$ for normal and grazing incidence geometries. Comprehensive transport data highlighted the Kondo effect in VSe$_{2.0}$ single crystals below 40 K \cite{Bar17}, thus, the small magnetic dichroism at the V \textit{L$_{2,3}$} edge can be assigned to Kondo impurities. Nevertheless, the small XMCD signal precludes us to retrieve a hysteresis loop from the magnetic dichroism experiments. Further, the introduction of Se vacancies seems to be detrimental to magnetism. 

In order to obtain a deep understanding of the electronic and magnetic ordering in single crystals of VSe$_{2.0}$, we have carried out \emph{ab initio} calculations, both in the normal (NS) and CDW state. ARPES and X-ray diffraction \cite{van_bruggen_magnetic_1976} have found a 3-dimensional CDW below 110 K, leading to the opening of a pseudo-gap at the Fermi level at ($\frac{1}{4}$, $\frac{1}{4}$, $\frac{1}{3}$) reciprocal lattice units \cite{terashima_charge-density_2003}. To take this into account in our calculations, we have computed a 4$a$ $\times$ 4$a$ $\times$ 3$c$ supercell. Therefore, the introduction of a periodic lattice distortion may affect the calculated DOS at the Fermi level and hence the magnetic properties of this itinerant electron system.

Since  VSe$_{2}$ is an itinerant-electron system, one can make use of the phenomenological Stoner model \cite{Stoner_crit} to analyze how close the system is to a magnetic instability. The Stoner theory compares the energy gained by the system via a spin splitting to the kinetic energy cost produced by displacing minority-spin electrons into a higher-energy majority-spin band. Only when the overall energy gets reduced, an itinerant electron system becomes spontaneously magnetic. The Stoner criterion states that the system will be FM if $I\cdot DOS(E_F)>1$ and non-magnetic otherwise, where $I$ is the exchange energy between the Bloch \textit{d}-band electrons, the so called Stoner parameter \cite{moriya1985spin}. Itinerant ferromagnets such as Fe, Ni and Co present values of $I\cdot DOS(E_F)$ ranging from 2.5 to 3 \cite{stoner_atomicnumber}.

\begin{figure}[!h]
  \centering
  \includegraphics[width=0.8\columnwidth]%
    {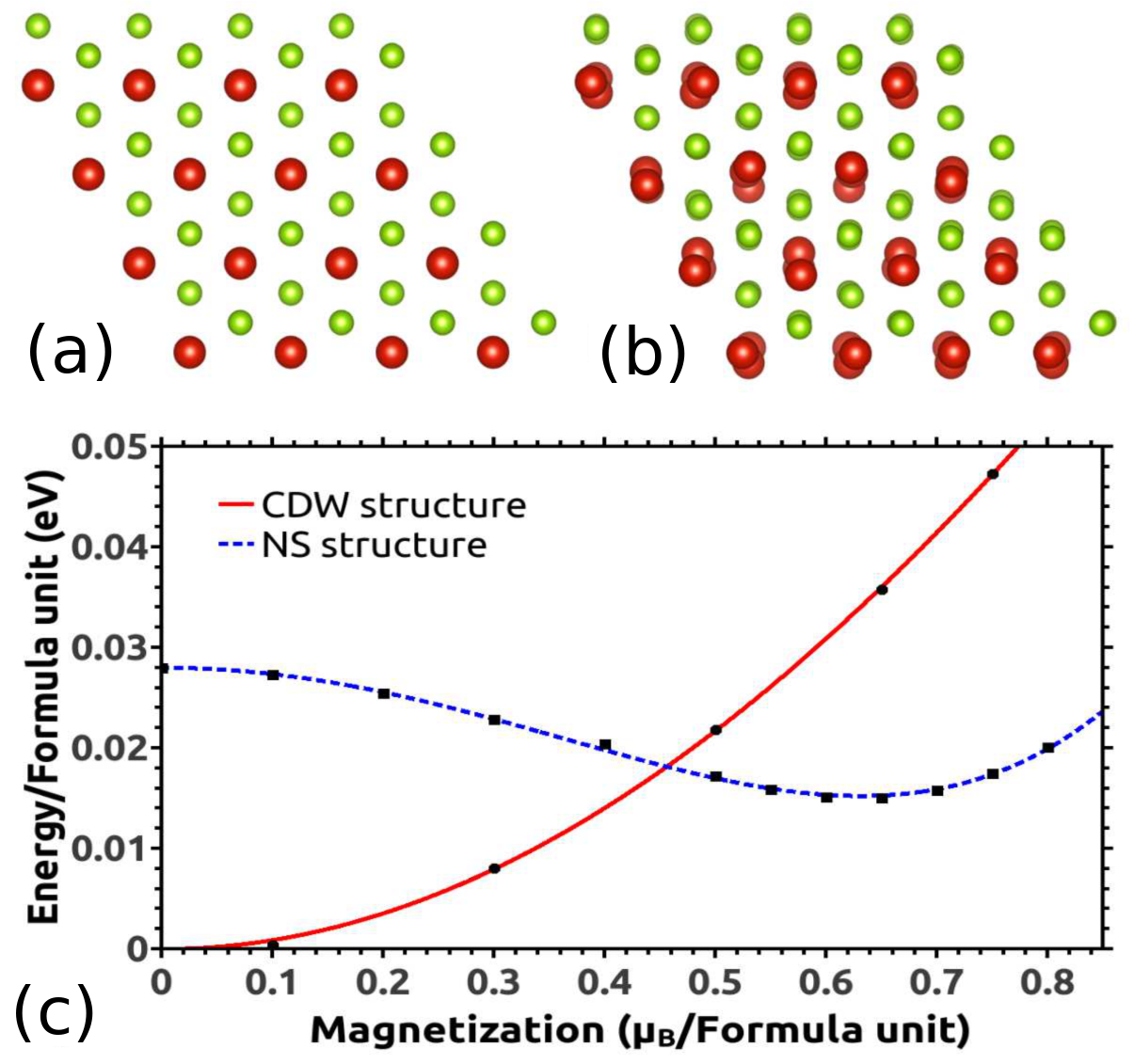}
     \caption{Top view of VSe$_{2.0}$ bulk structures. V (Se) atoms are represented as big red (green) spheres. a) NS structure in the \textit{P$\overline{3}$m1} space group. b) A modulated 4a $\times$ 4a $\times$ 3c supercell of the CDW state. c) Energy as a function of the magnetization for the bulk VSe$_{2.0}$ for the NS structure (blue line) and CDW state (red line). The minimum at 0.6 $\mu_B$ per V atom indicates that the magnetic solution is the most stable in the NS. At low temperature, the minimum-energy configuration is a non-magnetic CDW ground state.}\label{energetics}
\end{figure} 



\begin{figure}[!h]
  \centering
  \includegraphics[width=0.8\columnwidth]%
    {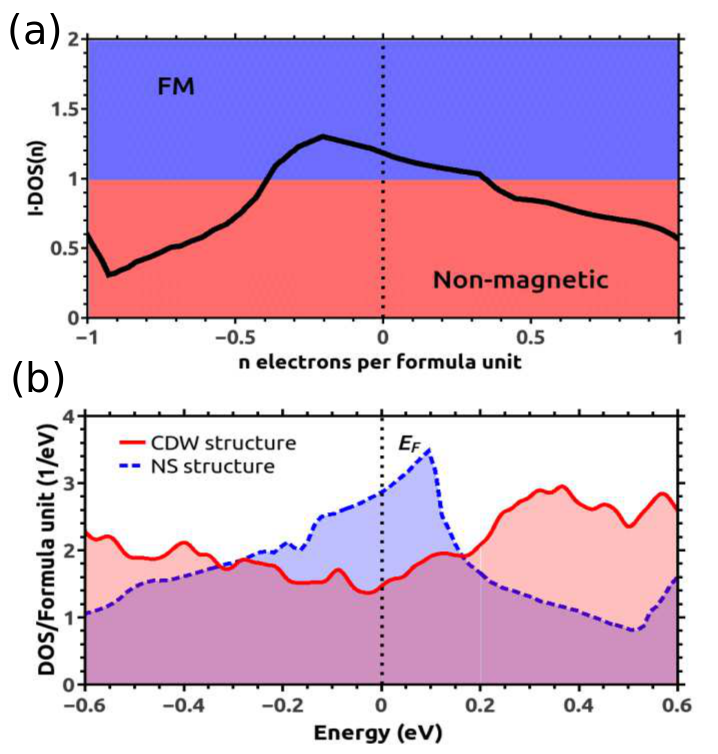}
    \caption{a) Stoner criterion for the NS structure as a function of the number of electrons introduced per formula unit. In the NS, VSe$_{2}$ is FM but close to a non-magnetic state. b) DOS around the Fermi level for the bulk structures. Red (blue), DOS of the CDW (NS) state, showing a reduction of the DOS in the CDW state and the breakdown of the magnetism in the charge ordered state.}\label{stoner_dos}
\end{figure}

It follows from the energy (E) \emph{vs} magnetization (M) curve that the bulk normal state yields a FM ground state with a broad minimum around at $0.6$ $\mu_B$ per V atom, (blue line in Fig. \ref{energetics}c), with $E=(1-I DOS(E_F))/DOS(E_F) M^2 +k M^4$ and $k$ is a fitting parameter independent of $I$ \cite{moriya1985spin}. The Fermi level is located at $n=0$, where \textit{n} is the number of electrons per formula unit and $n>0$ ($n<0$) implies hole (electron) doping, and the carrier concentration was calculated using a rigid-band approximation by integrating the total DOS of the non-magnetic calculation. In Fig. \ref{stoner_dos}a, we show that the Stoner criterion for FM is satisfied for -0.5$<$\textit{n}$<$0.5. However, any perturbation to this system leads to a small reduction in the DOS at the Fermi level and to a non-magnetic state. Previous \emph{ab initio} studies have shown that a reduction of the FM moment can be achieved by strain engineering \cite{ma_evidence_2012}. Very recently, W. Zhang et al. \cite{Zhang2019} took advantage of the proximity of VSe$_2$ to a magnetic ground state to engineer a FM heterostructure of VSe$_2$ with a magnetic moment of about 0.4 $\mu_B$ per V atom, as we have predicted here. This finding highlights that monolayers of VSe$_2$ can be manipulated to tailor new magnetic ground states. 


The optimized atomic positions at low temperature (CDW state), taking into account the new lattice periodicity \cite{van_bruggen_magnetic_1976}, are plotted in Fig. \ref{energetics}b, showing that the short-range hexagonal symmetry is lost. The CDW structure calculated \textit{ab-initio} is $28$ $meV$ per formula unit more stable than the NS structure resulting in a strong reduction of the DOS at the Fermi level (Fig. \ref{stoner_dos}b) and a quenching of the FM moment as compared with the NS (Fig. \ref{energetics}c).

\begin{figure}[!h]
  \centering
  \includegraphics[width=0.8\columnwidth]%
    {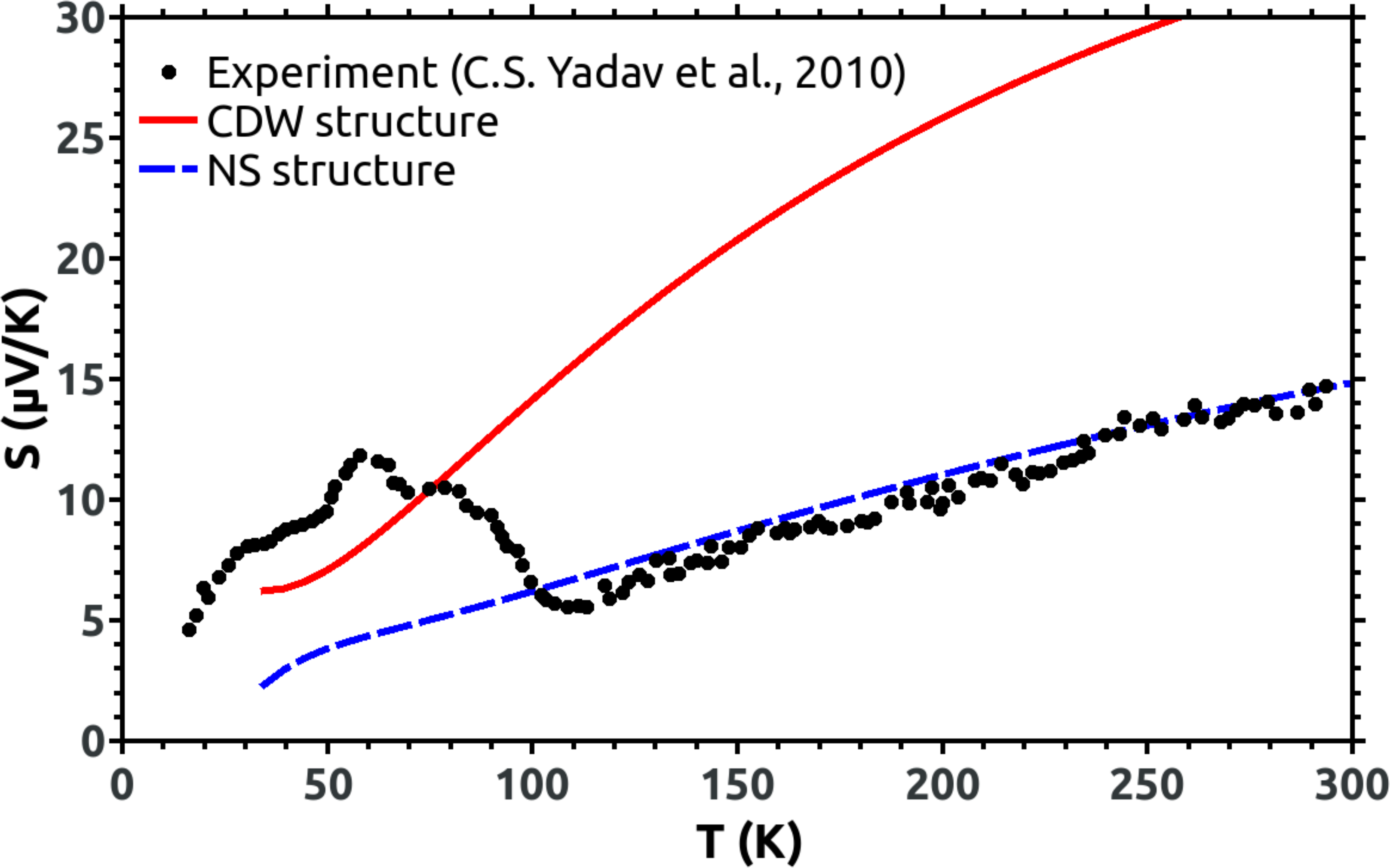}
    \caption{Thermopower as a function of temperature for bulk VSe$_2$. The black points show the experimental measurements from Ref. \cite{yadav_electronic_2010}. 
The red and blue lines show the calculated thermopower for the 4$a$ $\times$ 4$a$ $\times$ 3$c$ supercell (CDW state) and for the \textit{P$\overline{3}$m1} cell (normal state).
}
\label{seebeck}
\end{figure}

Experimentally, a significant enhancement in the Seebeck effect is observed at the transition from the NS at high temperatures to the CDW state below 110 K, due to the opening of a pseudo-gap at the Fermi level. In Fig. \ref{seebeck} the computed thermopower of both the NS (blue dashed line) and the CDW (red line) structures is compared to the ones in the literature \cite{yadav_electronic_2010} (black points), evidencing that the relaxed structure can be modeled reasonably well with the CDW state found experimentally. Despite the coarse fitting at low temperature, an enhancement of the thermopower with respect to the NS is also obtained in the theoretical simulations within the constant scattering time approximation. This is further evidence for the reliability of modelling bulk VSe$_{2}$ in its CDW phase using a 4$a$ $\times$ 4$a$ $\times$ 3$c$ supercell.

\subsection{\label{mono}Monolayer VSe$_2$}

\begin{figure}[!h]
  \centering
  \includegraphics[width=\columnwidth]%
    {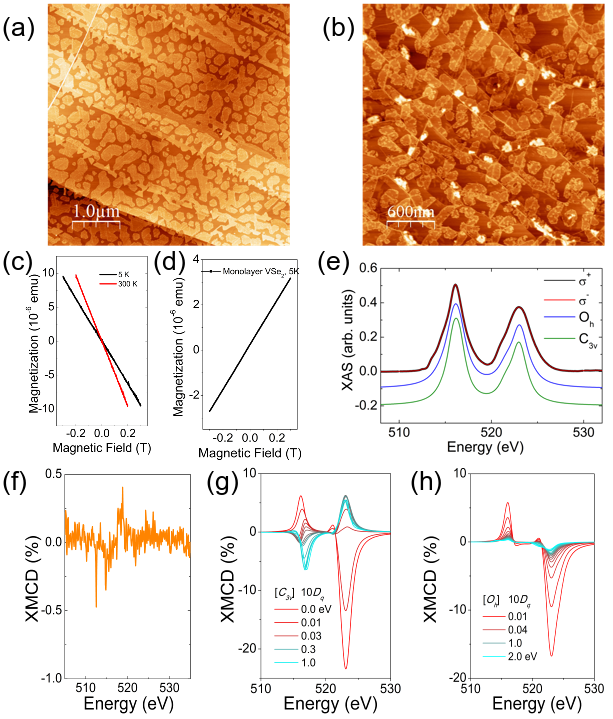}
     \caption{(a-b) AFM images of monolayer VSe$_{2.0}$ grown on HOPG and Si-C. (c) Magnetization vs field for VSe$_{2.0}$ on Si-C at 5 an 300 K. (d) Magnetization vs field for VSe$_{2.0}$ after the subtraction of the magnetic signal of the Si-C substrate. (e) Experimental XAS spectrum for ultrathin VSe$_2$ for $\sigma$$^+$ (black) and $\sigma$$^-$ (red) polarized light and the calculated isotropic XAS spectra for \textit{O}$_h$ and \textit{C}$_{3v}$ symmetries. (f) Experimental XMCD of VSe$_{2.0}$ on Si-C. (g-h) Calculated XMCD spectra for \textit{C}$_{3v}$ and \textit{O}$_h$ symmetries as a function of the crystal field splitting, 10\textit{D}$_q$.}\label{afm}
\end{figure}

Figure \ref{afm}(a-b) shows the atomic force microscopy (AFM) image of high quality monolayer VSe$_{2.0}$ grown on HOPG and Si-C, respectively, with typical heights of 6 \AA. The field dependence of the magnetization (\textit{M}[\textit{H}]) at 300 and 5 K for VSe$_{2.0}$ grown on Si-C substrate is presented in Figure \ref{afm}(c). The magnetic curves show a negative slope characteristic of the diamagnetic signal from the Si-C substrate, without indications of saturation or hysteresis, ruling out any magnetism coming from the VSe$_{2.0}$ monolayers, at least within the limit of detection of our setup. This is confirmed in \ref{afm}(d) after the subtraction of the magnetic signal of the Si-C substrate. Besides, XMCD at 6 T shows a featureless magnetic dichroism in normal or grazing incidence, as shown in Fig. \ref{afm}(f). Here, we point out that similar magnetic dichroism has been recently reported for spin frustrated monolayers of VSe$_2$\cite{Wong2019}. In addition, a careful comparison between the XAS spectra of the single crystals and the monolayer also reveals a broadening of the \textit{L}$_3$ edge in the ultra-thin limit, presumably as a consequence of the electronic reconstruction in the 2D limit.

To estimate the theoretical dichroism expected for a \textit{d}${^1}$ system, we have carried out cluster calculations in the octahedral and trigonal crystal field for the atomic-like 3\textit{d} transitions using  QUANTY. The code incorporates the intra-atomic 3\textit{d}-3\textit{d} and 2\textit{p}-3\textit{d}, magnetic exchange interactions, the atomic 2\textit{p} and 3\textit{d} spin-orbit couplings and local crystal field parameter and Coulomb energies (Slater integrals) obtained within the Hartree-Fock approximation \cite{Hav05}.

\begin{figure}[!h]
  \centering
  \includegraphics[width=0.8\columnwidth]%
    {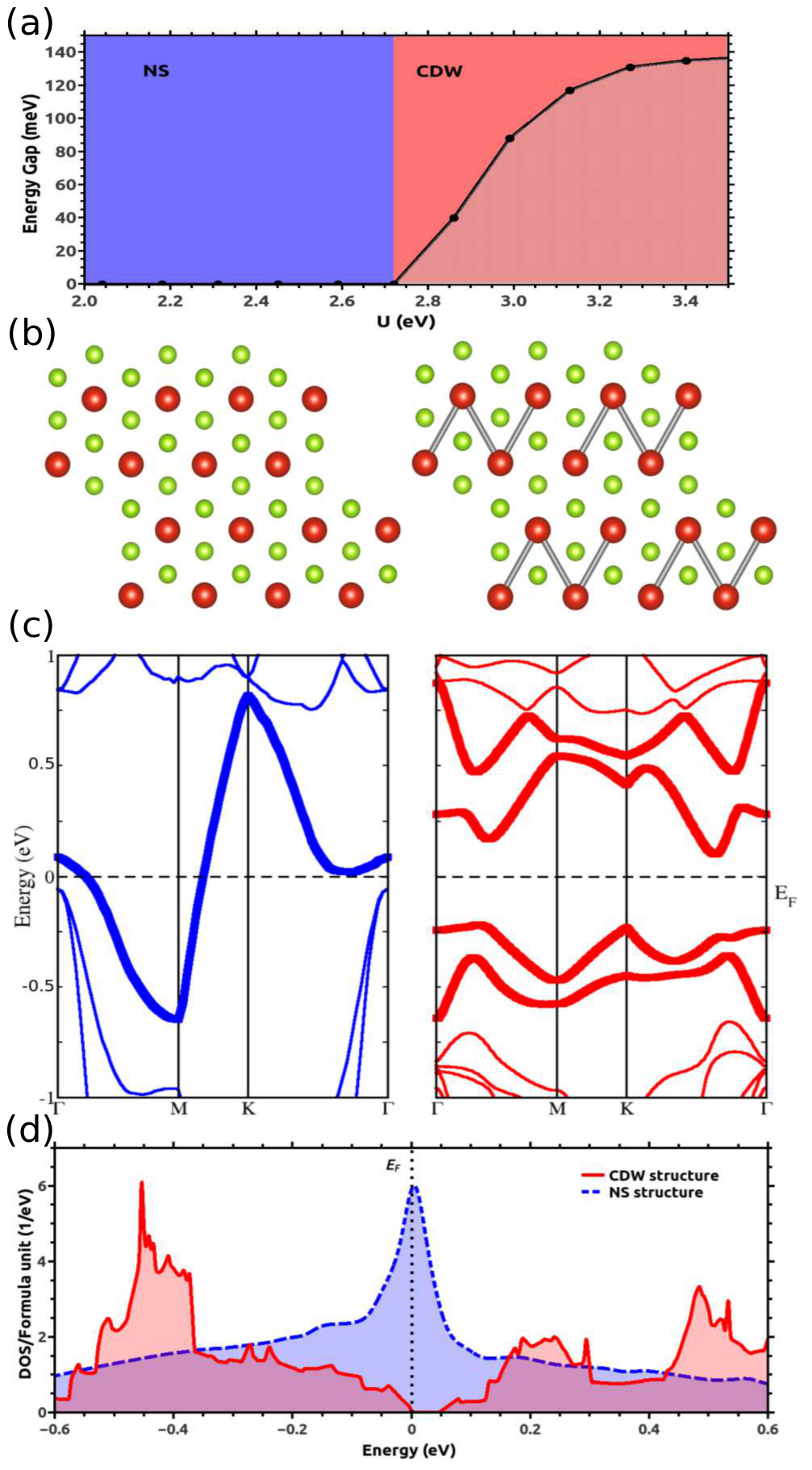}
\caption{Results for the monolayer. (a) Mapping on \textit{U}. The structure was fully optimized for each value of the Coulomb repulsion. At low values, the metallic FM NS structure is the most stable. For U$>2.7$ eV, a non-magnetic gap opens and the CDW stabilizes. (b) Structure schemes: NS on the left, each V atom has 6 neighbors at the same distance. 
CDW on the right, a tetramer is formed in a 2$a$ $\times$ 2$a$ supercell. The tetramer bonds are depicted in gray. 
(c) Band structures: On the left the NS, a d-band crosses the Fermi level producing a FM metallic state. On the right the CDW, the tetramer forms of two bonding and two antibonding bands, opening a gap and quenching the magnetic moment. (d) Comparison of the DOS of the NS monolayer VSe$_2$ (blue dashed line) with the CDW state (red line).}
\label{mono_calc}
\end{figure}

Figure \ref{afm}f and \ref{afm}(g-h) show the experimental and the simulated isotropic spectrum, defined as $\sigma$${^+}$+$\sigma$${^-}$ as a function of the crystal field 10\textit{D}$_q$ expected for spin-$\frac{1}{2}$ in a trigonal and octahedral symmetries. To simulate the monolayer VSe$_{2}$ spectra, we used 40\% of the atomic values of the Slater integrals in the \textit{O}$_h$ and \textit{C}$_{3v}$ symmetries of the V sites. As shown in Fig. \ref{afm} (g-h), the crystal field calculations for spin-$\frac{1}{2}$ V$^{4+}$ reveal a finite, but crystal-field dependent, 10\textit{D}$_q$, magnetic dichroism. Strain effects induced by the HOPG and SiC-graphene, which might alter the crystal field parameter, are found to be negligible in 2D monolayer TMDs. Therefore, the absence of magnetic dichroism in single layer VSe$_{2.0}$ points to a strong electronic renormalization at the ultrathin limit. 
In fact, screening effects are enhanced with respect to the bulk VSe$_{2.0}$, thus leading to an additional renormalization of the DOS. Indeed, the CDW phase is observed at \textit{T}$^{\mathrm{2D}}_{\mathrm{\textit{CDW}}}$ $>$ \textit{T}$^{\mathrm{3D}}_{\mathrm{\textit{CDW}}}$ \cite{xu_ultrathin_2013}, revealing an enhancement of the electron-electron and  electron-phonon interactions in 2D. Furthermore, the hopping parameter, \textit{t}, between layers vanishes and, hence, the ratio between the \textit{on-site} Coulomb repulsion and hopping, \textit{U}$/$\textit{t}, increases.

Following ARPES experiments \cite{nano_monolayer,PRL_monoVSe2,Coelho2019}, the ground state of monolayer VSe$_{2}$ is characterized by an energy gap. However, the electronic modulation of the CDW remains under discussion \cite{nano_monolayer,Umemoto2018,PRL_monoVSe2}. Therefore, our aim within the following DFT calculations will be to explain the physical mechanism that undergoes the monolayer rather than looking for perfect agreement with controversial experimental data. We present calculations on a 2$a$ $\times$ 2$a$ supercell since it was computationally affordable. This will allow us to understand the possible electronic reconstructions that may occur in the 2D limit when a periodic lattice distortion takes place, shedding light on how both the FM quenching and the full-gap opening occur. In order to include the electron interactions that become stronger in the monolayer limit, we have performed LDA+U calculations. Figure \ref{mono_calc}(a) shows the evolution of the energy gap in the whole Brillouin zone as a function of \textit{U} for a fully relaxed 2$a$ $\times$ 2$a$ supercell of the monolayer. At reduced values of \textit{U}, a metallic FM structure is stable. However, for \textit{U} greater than $2.7$ eV a non-magnetic CDW is formed and, consequently, an energy gap appears in the whole Brillouin zone, in agreement with experiments \cite{PRL_monoVSe2}. The calculated band structures for the non-magnetic-monolayer NS and CDW state are shown in Fig. \ref{mono_calc}(c). If no structural instability is present, the \textit{d}-band crossing at the Fermi level and the high density of states (blue dashed line in Fig. \ref{mono_calc}(d)) lead to a FM metallic state. Nevertheless, the structural transition associated to the CDW, depicted in Fig. \ref{mono_calc}(b), drives the system towards a Peierls-like distortion with a tetramerization among 4 V atoms. A comparison between the DOS of the NS structures both for the bulk and the monolayer reveals that decreasing dimensionality increases the DOS at the Fermi level, the bands become flatter due to the absence of the off-plane hopping. In general, for itinerant systems, this would be a mechanism to enhance the stability of a FM phase. However, due to the CDW state, the 4 \textit{d}-bands in the 2$a$ $\times$ 2$a$ supercell hybridize forming 2 bonding and 2 antibonding bands (right side of Fig. \ref{mono_calc}(b) opening an energy gap (red line in Fig. \ref{mono_calc}(d)) with a concomitant quenching of the FM moment.

\begin{figure}[!h]
  \centering
  \includegraphics[width=0.8\columnwidth]%
    {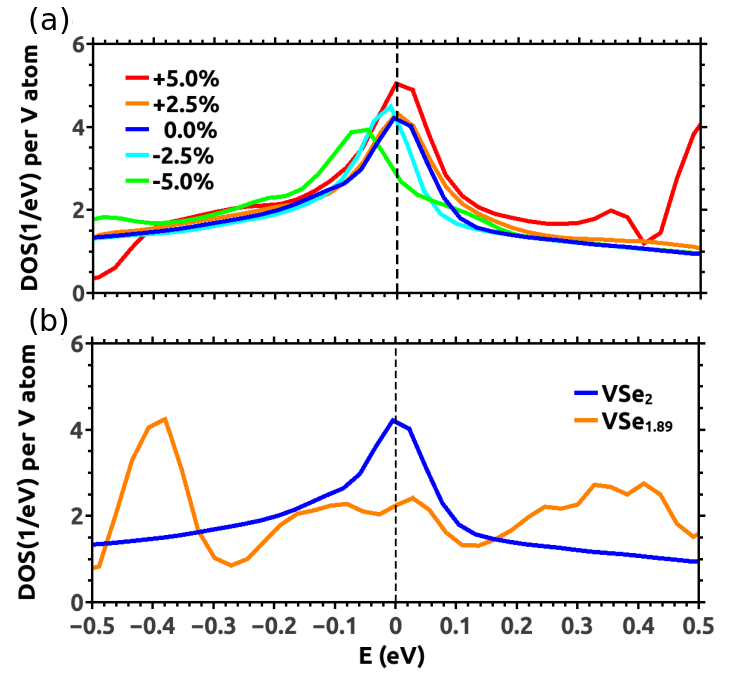}
     \caption{ Calculated DOS for the monolayer. (a) DOS for different strain values of the lattice parameter for the monolayer single-unit cell. A substantial change in the DOS at the Fermi level is not observed.
(b) Effect of Se vacancies on VSe$_{2}$ DOS. Including vacancies reduces the density of states at the Fermi level and hence decreases the possibility of having a FM phase.}
\label{dos_VSe18}
\end{figure}

Finally, in order to study \textit{ab initio} the intrinsic effect of strain and Se vacancies in the monolayer limit, we have computed the DOS for different strain values and also carried out similar calculations including Se vacancies in a V:Se ratio similar to our single-crystal experiments (Fig. \ref{dos_VSe18}). We have performed calculations in the NS and shown that the DOS is not substantially increased (mostly reduced) with respect to the unstrained-stoichiometric NS monolayer VSe$_{2.0}$. The effect of strain and Se vacancies can be seen in Fig. \ref{dos_VSe18}(a-b), suggesting that neither strain nor Se vacancies will induce an intrinsic FM state in single-layer VSe$_{2.0}$, as discussed above in our experiments for the bulk case.

\section{\label{conclusions}DISCUSSION AND CONCLUSIONS}

Recently, transport and magnetometry experiments have reported a FM ground state for monolayers of VSe$_2$ grown on different substrates \cite{bonilla_strong_2018}. This is in agreement with previous DFT calculations that have predicted the emergence of a FM ground state for monolayers of VX$_2$ (X=S, Se) \cite{li_versatile_2014, ma_evidence_2012, PhysRevB_DI, NJP}. Our combination of experimental data, including magnetization and XMCD, find a small ferromagnetic signal for stoichiometric bulk VSe$_{2.0}$, which may arise from V impurities intercalated in the van der Waals gap. Naively, one may think that this Kondo mechanism gets inoperative when reducing the dimensionality of VSe$_{2.0}$ down to the ultrathin limit, since no van der Waals gap is present. Nonetheless, still impurities in the surface of the monolayer can produce a Kondo effect. In the bulk case, \textit{ab initio} calculations show that a CDW phase of the periodicity found experimentally destroys magnetism and is the ground state structure. Recently, first principles calculations of the CDW structure of monolayer VSe$_2$ also reported a distorted CDW structure more stable than the FM one, indicating a competition of magnetism and CDW \cite{Coelho2019}. 

Following our XMCD data and DFT calculations, we have found a non-magnetic ground state in monolayer VSe$_2$, in agreement with recent ARPES data \cite{nano_monolayer,PRL_monoVSe2} that demonstrates the absence of spin-polarized bands in the monolayer limit \cite{PRL_monoVSe2}. More recent work \cite{Wong2019} also reports a non-magnetic but frustrated ground state in monolayer VSe$_2$ due to a competition  between ferro and antiferromagnetic orders. In addition, defects are shown to be detrimental to ferromagnetism, unlike  the emergence of the vacancy-induced magnetism observed in graphite \cite{Cer09}. Although our magnetization data does not allow us to reveal a frustrated magnetic ground state, the absence of long range magnetic order due to disorder and vacancies corroborates such magnetic frustration. 
Nevertheless, despite our DFT calculations find a drop of DOS at the Fermi level discarding a possible FM state, the effect of intrinsic disorder in monolayer VSe$_{2}$ has to be carefully studied experimentally, since high density of vacancies, anti-sites, substitutions, edges and grain boundaries can dominate the materials\textsc{\char13} properties. 

In conclusion, we have studied experimentally and theoretically the ground state of bulk and monolayers of VSe$_{2}$. Our DFT calculations predict that the ground state of bulk VSe$_2$ develops a commensurate lattice distortion with a 4$a$ $\times$ 4$a$ $\times$ 3$c$ supercell that reproduces the variation of the thermopower at high and low temperature,  demonstrating the importance of considering the correct ground state structure when performing \emph{ab initio}  studies in magnetic materials. This structure is shown to be related with the CDW phase that has been experimentally detected at low temperatures. Our DFT calculations for such a ground state show that FM is destroyed by such distortion, VSe$_{2}$ being a paramagnet in the bulk but on the verge to a magnetic state. This result confirms previous analysis that highlight the importance of phonon instabilities, besides electronic ones, in driving the formation of the CDW state \cite{PhysRevB.77.165135}. It also reconciles theory and experiment. In the monolayer limit, we found that a periodic lattice distortion (2$a$ $\times$ 2$a$) associated to the CDW state is sufficient to open an energy gap through a Peierls-like distortion destroying the tendency towards a FM state, as we have reported experimentally by means of magnetization and XMCD. 
Our calculations suggest that the origin of the magnetic signal obtained for VSe$_{2.0}$ cannot be related to the band structure of the material, either in the bulk or in the single-layer limit and any magnetic ground state in the monolayer limit might be associated to a high density of defects, edges or grain boundaries, although control experiments in bulk VSe$_{1.8}$ and \textit{ab initio} calculations in single layer VSe$_{2.0}$ indicate that strain induced by vacancies work against ferromagnetism in clear contrast to the magnetic ground state recently reported in few layer PtSe$_2$ \cite{Avsar2019}, presumably due to the strong competition between ferro- and antiferromagnetic states \cite{Wong2019}.

\section{Acknowledgements}

This work is supported by the MINECO of Spain through the project MAT2016-80762-R, PGC2018-101334-A-C22 and PGC2018-101334-B-C21. A.O.F. thanks MECD for the financial support received through the FPU grant FPU16/02572. M.G. acknowledges funding from the European Commission through the Marie Sklodowska-Curie IEF project SUPER2D (GA-748971).
M.M.U. acknowledges support by the Spanish MINECO under grant no.  MAT2017-82074-ERC and by the ERC Starting grant “LINKSPM” (Grant 758558). A. Berger and L. H. Hueso are acknowledged for sharing CIC nanoGUNE facilities and the SQUID measurements. S.B-C thanks IKERBASQUE for financial support. We also thank J. Fern\'{a}ndez-Rossier, I. Oleynik, Warren E. Pickett and D. Soriano for fruitful discussions and ALBA Synchrotron Light Facility for the provision of synchrotron beamtime.


\begin{thebibliography}{62}%
\makeatletter
\providecommand \@ifxundefined [1]{%
 \@ifx{#1\undefined}
}%
\providecommand \@ifnum [1]{%
 \ifnum #1\expandafter \@firstoftwo
 \else \expandafter \@secondoftwo
 \fi
}%
\providecommand \@ifx [1]{%
 \ifx #1\expandafter \@firstoftwo
 \else \expandafter \@secondoftwo
 \fi
}%
\providecommand \natexlab [1]{#1}%
\providecommand \enquote  [1]{``#1''}%
\providecommand \bibnamefont  [1]{#1}%
\providecommand \bibfnamefont [1]{#1}%
\providecommand \citenamefont [1]{#1}%
\providecommand \href@noop [0]{\@secondoftwo}%
\providecommand \href [0]{\begingroup \@sanitize@url \@href}%
\providecommand \@href[1]{\@@startlink{#1}\@@href}%
\providecommand \@@href[1]{\endgroup#1\@@endlink}%
\providecommand \@sanitize@url [0]{\catcode `\\12\catcode `\$12\catcode
  `\&12\catcode `\#12\catcode `\^12\catcode `\_12\catcode `\%12\relax}%
\providecommand \@@startlink[1]{}%
\providecommand \@@endlink[0]{}%
\providecommand \url  [0]{\begingroup\@sanitize@url \@url }%
\providecommand \@url [1]{\endgroup\@href {#1}{\urlprefix }}%
\providecommand \urlprefix  [0]{URL }%
\providecommand \Eprint [0]{\href }%
\providecommand \doibase [0]{http://dx.doi.org/}%
\providecommand \selectlanguage [0]{\@gobble}%
\providecommand \bibinfo  [0]{\@secondoftwo}%
\providecommand \bibfield  [0]{\@secondoftwo}%
\providecommand \translation [1]{[#1]}%
\providecommand \BibitemOpen [0]{}%
\providecommand \bibitemStop [0]{}%
\providecommand \bibitemNoStop [0]{.\EOS\space}%
\providecommand \EOS [0]{\spacefactor3000\relax}%
\providecommand \BibitemShut  [1]{\csname bibitem#1\endcsname}%
\let\auto@bib@innerbib\@empty
\bibitem [{\citenamefont {Novoselov}\ \emph {et~al.}(2004)\citenamefont
  {Novoselov}, \citenamefont {Geim}, \citenamefont {Morozov}, \citenamefont
  {Jiang}, \citenamefont {Zhang}, \citenamefont {Dubonos}, \citenamefont
  {Grigorieva},\ and\ \citenamefont {Firsov}}]{Graphene}%
  \BibitemOpen
  \bibfield  {author} {\bibinfo {author} {\bibfnamefont {K.~S.}\ \bibnamefont
  {Novoselov}}, \bibinfo {author} {\bibfnamefont {A.~K.}\ \bibnamefont {Geim}},
  \bibinfo {author} {\bibfnamefont {S.~V.}\ \bibnamefont {Morozov}}, \bibinfo
  {author} {\bibfnamefont {D.}~\bibnamefont {Jiang}}, \bibinfo {author}
  {\bibfnamefont {Y.}~\bibnamefont {Zhang}}, \bibinfo {author} {\bibfnamefont
  {S.~V.}\ \bibnamefont {Dubonos}}, \bibinfo {author} {\bibfnamefont {I.~V.}\
  \bibnamefont {Grigorieva}}, \ and\ \bibinfo {author} {\bibfnamefont {A.~A.}\
  \bibnamefont {Firsov}},\ }\href {\doibase 10.1126/science.1102896} {\bibfield
   {journal} {\bibinfo  {journal} {Science}\ }\textbf {\bibinfo {volume}
  {306}},\ \bibinfo {pages} {666} (\bibinfo {year} {2004})},\ \Eprint
  {http://arxiv.org/abs/http://science.sciencemag.org/content/306/5696/666.full.pdf}
  {http://science.sciencemag.org/content/306/5696/666.full.pdf} \BibitemShut
  {NoStop}%
\bibitem [{\citenamefont {Manzeli}\ \emph {et~al.}(2017)\citenamefont
  {Manzeli}, \citenamefont {Ovchinnikov}, \citenamefont {Pasquier},
  \citenamefont {Yazyev},\ and\ \citenamefont {Kis}}]{manzeli_2d_2017}%
  \BibitemOpen
  \bibfield  {author} {\bibinfo {author} {\bibfnamefont {S.}~\bibnamefont
  {Manzeli}}, \bibinfo {author} {\bibfnamefont {D.}~\bibnamefont
  {Ovchinnikov}}, \bibinfo {author} {\bibfnamefont {D.}~\bibnamefont
  {Pasquier}}, \bibinfo {author} {\bibfnamefont {O.~V.}\ \bibnamefont
  {Yazyev}}, \ and\ \bibinfo {author} {\bibfnamefont {A.}~\bibnamefont {Kis}},\
  }\href {\doibase 10.1038/natrevmats.2017.33} {\bibfield  {journal} {\bibinfo
  {journal} {Nature Reviews Materials}\ }\textbf {\bibinfo {volume} {2}},\
  \bibinfo {pages} {17033} (\bibinfo {year} {2017})}\BibitemShut {NoStop}%
\bibitem [{\citenamefont {Choi}\ \emph {et~al.}(2017)\citenamefont {Choi},
  \citenamefont {Choudhary}, \citenamefont {Han}, \citenamefont {Park},
  \citenamefont {Akinwande},\ and\ \citenamefont {Lee}}]{CHOI2017116}%
  \BibitemOpen
  \bibfield  {author} {\bibinfo {author} {\bibfnamefont {W.}~\bibnamefont
  {Choi}}, \bibinfo {author} {\bibfnamefont {N.}~\bibnamefont {Choudhary}},
  \bibinfo {author} {\bibfnamefont {G.~H.}\ \bibnamefont {Han}}, \bibinfo
  {author} {\bibfnamefont {J.}~\bibnamefont {Park}}, \bibinfo {author}
  {\bibfnamefont {D.}~\bibnamefont {Akinwande}}, \ and\ \bibinfo {author}
  {\bibfnamefont {Y.~H.}\ \bibnamefont {Lee}},\ }\href {\doibase
  https://doi.org/10.1016/j.mattod.2016.10.002} {\bibfield  {journal} {\bibinfo
   {journal} {Materials Today}\ }\textbf {\bibinfo {volume} {20}},\ \bibinfo
  {pages} {116 } (\bibinfo {year} {2017})}\BibitemShut {NoStop}%
\bibitem [{\citenamefont {Zhang}\ and\ \citenamefont
  {Zhang}(2017)}]{C7TC01088E}%
  \BibitemOpen
  \bibfield  {author} {\bibinfo {author} {\bibfnamefont {G.}~\bibnamefont
  {Zhang}}\ and\ \bibinfo {author} {\bibfnamefont {Y.-W.}\ \bibnamefont
  {Zhang}},\ }\href {\doibase 10.1039/C7TC01088E} {\bibfield  {journal}
  {\bibinfo  {journal} {J. Mater. Chem. C}\ }\textbf {\bibinfo {volume} {5}},\
  \bibinfo {pages} {7684} (\bibinfo {year} {2017})}\BibitemShut {NoStop}%
\bibitem [{\citenamefont {Xia}\ \emph {et~al.}(2017)\citenamefont {Xia},
  \citenamefont {Yan},\ and\ \citenamefont {Shen}}]{XIA20171}%
  \BibitemOpen
  \bibfield  {author} {\bibinfo {author} {\bibfnamefont {J.}~\bibnamefont
  {Xia}}, \bibinfo {author} {\bibfnamefont {J.}~\bibnamefont {Yan}}, \ and\
  \bibinfo {author} {\bibfnamefont {Z.~X.}\ \bibnamefont {Shen}},\ }\href
  {\doibase https://doi.org/10.1016/j.flatc.2017.06.007} {\bibfield  {journal}
  {\bibinfo  {journal} {FlatChem}\ }\textbf {\bibinfo {volume} {4}},\ \bibinfo
  {pages} {1 } (\bibinfo {year} {2017})}\BibitemShut {NoStop}%
\bibitem [{\citenamefont {Akinwande}\ \emph {et~al.}(2014)\citenamefont
  {Akinwande}, \citenamefont {Petrone},\ and\ \citenamefont {Hone}}]{Aki14}%
  \BibitemOpen
  \bibfield  {author} {\bibinfo {author} {\bibfnamefont {D.}~\bibnamefont
  {Akinwande}}, \bibinfo {author} {\bibfnamefont {N.}~\bibnamefont {Petrone}},
  \ and\ \bibinfo {author} {\bibfnamefont {J.}~\bibnamefont {Hone}},\ }\href
  {https://doi.org/10.1038/ncomms6678} {\bibfield  {journal} {\bibinfo
  {journal} {Nature Communications}\ }\textbf {\bibinfo {volume} {5}},\
  \bibinfo {pages} {5678 EP } (\bibinfo {year} {2014})},\ \bibinfo {note}
  {review Article}\BibitemShut {NoStop}%
\bibitem [{\citenamefont {Xu}\ \emph {et~al.}(2013)\citenamefont {Xu},
  \citenamefont {Chen}, \citenamefont {Li}, \citenamefont {Wu}, \citenamefont
  {Guo}, \citenamefont {Zhao}, \citenamefont {Wu},\ and\ \citenamefont
  {Xie}}]{xu_ultrathin_2013}%
  \BibitemOpen
  \bibfield  {author} {\bibinfo {author} {\bibfnamefont {K.}~\bibnamefont
  {Xu}}, \bibinfo {author} {\bibfnamefont {P.}~\bibnamefont {Chen}}, \bibinfo
  {author} {\bibfnamefont {X.}~\bibnamefont {Li}}, \bibinfo {author}
  {\bibfnamefont {C.}~\bibnamefont {Wu}}, \bibinfo {author} {\bibfnamefont
  {Y.}~\bibnamefont {Guo}}, \bibinfo {author} {\bibfnamefont {J.}~\bibnamefont
  {Zhao}}, \bibinfo {author} {\bibfnamefont {X.}~\bibnamefont {Wu}}, \ and\
  \bibinfo {author} {\bibfnamefont {Y.}~\bibnamefont {Xie}},\ }\href {\doibase
  10.1002/anie.201304337} {\bibfield  {journal} {\bibinfo  {journal}
  {Angewandte Chemie International Edition}\ }\textbf {\bibinfo {volume}
  {52}},\ \bibinfo {pages} {10477} (\bibinfo {year} {2013})}\BibitemShut
  {NoStop}%
\bibitem [{\citenamefont {Feng}\ \emph {et~al.}(2018)\citenamefont {Feng},
  \citenamefont {Biswas}, \citenamefont {Rajan}, \citenamefont {Watson},
  \citenamefont {Mazzola}, \citenamefont {Clark}, \citenamefont {Underwood},
  \citenamefont {Marković}, \citenamefont {McLaren}, \citenamefont {Hunter},
  \citenamefont {Burn}, \citenamefont {Duffy}, \citenamefont {Barua},
  \citenamefont {Balakrishnan}, \citenamefont {Bertran}, \citenamefont
  {Le~Fèvre}, \citenamefont {Kim}, \citenamefont {van~der Laan}, \citenamefont
  {Hesjedal}, \citenamefont {Wahl},\ and\ \citenamefont
  {King}}]{nano_monolayer}%
  \BibitemOpen
  \bibfield  {author} {\bibinfo {author} {\bibfnamefont {J.}~\bibnamefont
  {Feng}}, \bibinfo {author} {\bibfnamefont {D.}~\bibnamefont {Biswas}},
  \bibinfo {author} {\bibfnamefont {A.}~\bibnamefont {Rajan}}, \bibinfo
  {author} {\bibfnamefont {M.~D.}\ \bibnamefont {Watson}}, \bibinfo {author}
  {\bibfnamefont {F.}~\bibnamefont {Mazzola}}, \bibinfo {author} {\bibfnamefont
  {O.~J.}\ \bibnamefont {Clark}}, \bibinfo {author} {\bibfnamefont
  {K.}~\bibnamefont {Underwood}}, \bibinfo {author} {\bibfnamefont
  {I.}~\bibnamefont {Marković}}, \bibinfo {author} {\bibfnamefont
  {M.}~\bibnamefont {McLaren}}, \bibinfo {author} {\bibfnamefont
  {A.}~\bibnamefont {Hunter}}, \bibinfo {author} {\bibfnamefont {D.~M.}\
  \bibnamefont {Burn}}, \bibinfo {author} {\bibfnamefont {L.~B.}\ \bibnamefont
  {Duffy}}, \bibinfo {author} {\bibfnamefont {S.}~\bibnamefont {Barua}},
  \bibinfo {author} {\bibfnamefont {G.}~\bibnamefont {Balakrishnan}}, \bibinfo
  {author} {\bibfnamefont {F.}~\bibnamefont {Bertran}}, \bibinfo {author}
  {\bibfnamefont {P.}~\bibnamefont {Le~Fèvre}}, \bibinfo {author}
  {\bibfnamefont {T.~K.}\ \bibnamefont {Kim}}, \bibinfo {author} {\bibfnamefont
  {G.}~\bibnamefont {van~der Laan}}, \bibinfo {author} {\bibfnamefont
  {T.}~\bibnamefont {Hesjedal}}, \bibinfo {author} {\bibfnamefont
  {P.}~\bibnamefont {Wahl}}, \ and\ \bibinfo {author} {\bibfnamefont
  {P.~D.~C.}\ \bibnamefont {King}},\ }\href {\doibase
  10.1021/acs.nanolett.8b01649} {\bibfield  {journal} {\bibinfo  {journal}
  {Nano Letters}\ }\textbf {\bibinfo {volume} {18}},\ \bibinfo {pages} {4493}
  (\bibinfo {year} {2018})},\ \bibinfo {note} {pMID: 29912565}\BibitemShut
  {NoStop}%
\bibitem [{\citenamefont {Umemoto}\ \emph {et~al.}(2019)\citenamefont
  {Umemoto}, \citenamefont {Sugawara}, \citenamefont {Nakata}, \citenamefont
  {Takahashi},\ and\ \citenamefont {Sato}}]{Umemoto2018}%
  \BibitemOpen
  \bibfield  {author} {\bibinfo {author} {\bibfnamefont {Y.}~\bibnamefont
  {Umemoto}}, \bibinfo {author} {\bibfnamefont {K.}~\bibnamefont {Sugawara}},
  \bibinfo {author} {\bibfnamefont {Y.}~\bibnamefont {Nakata}}, \bibinfo
  {author} {\bibfnamefont {T.}~\bibnamefont {Takahashi}}, \ and\ \bibinfo
  {author} {\bibfnamefont {T.}~\bibnamefont {Sato}},\ }\href {\doibase
  10.1007/s12274-018-2196-4} {\bibfield  {journal} {\bibinfo  {journal} {Nano
  Research}\ }\textbf {\bibinfo {volume} {12}},\ \bibinfo {pages} {165}
  (\bibinfo {year} {2019})}\BibitemShut {NoStop}%
\bibitem [{\citenamefont {Chen}\ \emph {et~al.}(2018)\citenamefont {Chen},
  \citenamefont {Pai}, \citenamefont {Chan}, \citenamefont {Madhavan},
  \citenamefont {Chou}, \citenamefont {Mo}, \citenamefont {Fedorov},\ and\
  \citenamefont {Chiang}}]{PRL_monoVSe2}%
  \BibitemOpen
  \bibfield  {author} {\bibinfo {author} {\bibfnamefont {P.}~\bibnamefont
  {Chen}}, \bibinfo {author} {\bibfnamefont {W.~W.}\ \bibnamefont {Pai}},
  \bibinfo {author} {\bibfnamefont {Y.-H.}\ \bibnamefont {Chan}}, \bibinfo
  {author} {\bibfnamefont {V.}~\bibnamefont {Madhavan}}, \bibinfo {author}
  {\bibfnamefont {M.~Y.}\ \bibnamefont {Chou}}, \bibinfo {author}
  {\bibfnamefont {S.-K.}\ \bibnamefont {Mo}}, \bibinfo {author} {\bibfnamefont
  {A.-V.}\ \bibnamefont {Fedorov}}, \ and\ \bibinfo {author} {\bibfnamefont
  {T.-C.}\ \bibnamefont {Chiang}},\ }\href {\doibase
  10.1103/PhysRevLett.121.196402} {\bibfield  {journal} {\bibinfo  {journal}
  {Phys. Rev. Lett.}\ }\textbf {\bibinfo {volume} {121}},\ \bibinfo {pages}
  {196402} (\bibinfo {year} {2018})}\BibitemShut {NoStop}%
\bibitem [{\citenamefont {Ugeda}\ \emph {et~al.}(2015)\citenamefont {Ugeda},
  \citenamefont {Bradley}, \citenamefont {Zhang}, \citenamefont {Onishi},
  \citenamefont {Chen}, \citenamefont {Ruan}, \citenamefont
  {Ojeda-Aristizabal}, \citenamefont {Ryu}, \citenamefont {Edmonds},
  \citenamefont {Tsai}, \citenamefont {Riss}, \citenamefont {Mo}, \citenamefont
  {Lee}, \citenamefont {Zettl}, \citenamefont {Hussain}, \citenamefont {Shen},\
  and\ \citenamefont {Crommie}}]{Uge16}%
  \BibitemOpen
  \bibfield  {author} {\bibinfo {author} {\bibfnamefont {M.~M.}\ \bibnamefont
  {Ugeda}}, \bibinfo {author} {\bibfnamefont {A.~J.}\ \bibnamefont {Bradley}},
  \bibinfo {author} {\bibfnamefont {Y.}~\bibnamefont {Zhang}}, \bibinfo
  {author} {\bibfnamefont {S.}~\bibnamefont {Onishi}}, \bibinfo {author}
  {\bibfnamefont {Y.}~\bibnamefont {Chen}}, \bibinfo {author} {\bibfnamefont
  {W.}~\bibnamefont {Ruan}}, \bibinfo {author} {\bibfnamefont {C.}~\bibnamefont
  {Ojeda-Aristizabal}}, \bibinfo {author} {\bibfnamefont {H.}~\bibnamefont
  {Ryu}}, \bibinfo {author} {\bibfnamefont {M.~T.}\ \bibnamefont {Edmonds}},
  \bibinfo {author} {\bibfnamefont {H.-Z.}\ \bibnamefont {Tsai}}, \bibinfo
  {author} {\bibfnamefont {A.}~\bibnamefont {Riss}}, \bibinfo {author}
  {\bibfnamefont {S.-K.}\ \bibnamefont {Mo}}, \bibinfo {author} {\bibfnamefont
  {D.}~\bibnamefont {Lee}}, \bibinfo {author} {\bibfnamefont {A.}~\bibnamefont
  {Zettl}}, \bibinfo {author} {\bibfnamefont {Z.}~\bibnamefont {Hussain}},
  \bibinfo {author} {\bibfnamefont {Z.-X.}\ \bibnamefont {Shen}}, \ and\
  \bibinfo {author} {\bibfnamefont {M.~F.}\ \bibnamefont {Crommie}},\ }\href
  {https://doi.org/10.1038/nphys3527} {\bibfield  {journal} {\bibinfo
  {journal} {Nature Physics}\ }\textbf {\bibinfo {volume} {12}},\ \bibinfo
  {pages} {92 EP } (\bibinfo {year} {2015})},\ \bibinfo {note}
  {article}\BibitemShut {NoStop}%
\bibitem [{\citenamefont {Tran}\ \emph {et~al.}(2019)\citenamefont {Tran},
  \citenamefont {Moody}, \citenamefont {Wu}, \citenamefont {Lu}, \citenamefont
  {Choi}, \citenamefont {Kim}, \citenamefont {Rai}, \citenamefont {Sanchez},
  \citenamefont {Quan}, \citenamefont {Singh}, \citenamefont {Embley},
  \citenamefont {Zepeda}, \citenamefont {Campbell}, \citenamefont {Autry},
  \citenamefont {Taniguchi}, \citenamefont {Watanabe}, \citenamefont {Lu},
  \citenamefont {Banerjee}, \citenamefont {Silverman}, \citenamefont {Kim},
  \citenamefont {Tutuc}, \citenamefont {Yang}, \citenamefont {MacDonald},\ and\
  \citenamefont {Li}}]{Tra19}%
  \BibitemOpen
  \bibfield  {author} {\bibinfo {author} {\bibfnamefont {K.}~\bibnamefont
  {Tran}}, \bibinfo {author} {\bibfnamefont {G.}~\bibnamefont {Moody}},
  \bibinfo {author} {\bibfnamefont {F.}~\bibnamefont {Wu}}, \bibinfo {author}
  {\bibfnamefont {X.}~\bibnamefont {Lu}}, \bibinfo {author} {\bibfnamefont
  {J.}~\bibnamefont {Choi}}, \bibinfo {author} {\bibfnamefont {K.}~\bibnamefont
  {Kim}}, \bibinfo {author} {\bibfnamefont {A.}~\bibnamefont {Rai}}, \bibinfo
  {author} {\bibfnamefont {D.~A.}\ \bibnamefont {Sanchez}}, \bibinfo {author}
  {\bibfnamefont {J.}~\bibnamefont {Quan}}, \bibinfo {author} {\bibfnamefont
  {A.}~\bibnamefont {Singh}}, \bibinfo {author} {\bibfnamefont
  {J.}~\bibnamefont {Embley}}, \bibinfo {author} {\bibfnamefont
  {A.}~\bibnamefont {Zepeda}}, \bibinfo {author} {\bibfnamefont
  {M.}~\bibnamefont {Campbell}}, \bibinfo {author} {\bibfnamefont
  {T.}~\bibnamefont {Autry}}, \bibinfo {author} {\bibfnamefont
  {T.}~\bibnamefont {Taniguchi}}, \bibinfo {author} {\bibfnamefont
  {K.}~\bibnamefont {Watanabe}}, \bibinfo {author} {\bibfnamefont
  {N.}~\bibnamefont {Lu}}, \bibinfo {author} {\bibfnamefont {S.~K.}\
  \bibnamefont {Banerjee}}, \bibinfo {author} {\bibfnamefont {K.~L.}\
  \bibnamefont {Silverman}}, \bibinfo {author} {\bibfnamefont {S.}~\bibnamefont
  {Kim}}, \bibinfo {author} {\bibfnamefont {E.}~\bibnamefont {Tutuc}}, \bibinfo
  {author} {\bibfnamefont {L.}~\bibnamefont {Yang}}, \bibinfo {author}
  {\bibfnamefont {A.~H.}\ \bibnamefont {MacDonald}}, \ and\ \bibinfo {author}
  {\bibfnamefont {X.}~\bibnamefont {Li}},\ }\href {\doibase
  10.1038/s41586-019-0975-z} {\bibfield  {journal} {\bibinfo  {journal}
  {Nature}\ }\textbf {\bibinfo {volume} {567}},\ \bibinfo {pages} {71}
  (\bibinfo {year} {2019})}\BibitemShut {NoStop}%
\bibitem [{\citenamefont {Seyler}\ \emph {et~al.}(2019)\citenamefont {Seyler},
  \citenamefont {Rivera}, \citenamefont {Yu}, \citenamefont {Wilson},
  \citenamefont {Ray}, \citenamefont {Mandrus}, \citenamefont {Yan},
  \citenamefont {Yao},\ and\ \citenamefont {Xu}}]{Sey19}%
  \BibitemOpen
  \bibfield  {author} {\bibinfo {author} {\bibfnamefont {K.~L.}\ \bibnamefont
  {Seyler}}, \bibinfo {author} {\bibfnamefont {P.}~\bibnamefont {Rivera}},
  \bibinfo {author} {\bibfnamefont {H.}~\bibnamefont {Yu}}, \bibinfo {author}
  {\bibfnamefont {N.~P.}\ \bibnamefont {Wilson}}, \bibinfo {author}
  {\bibfnamefont {E.~L.}\ \bibnamefont {Ray}}, \bibinfo {author} {\bibfnamefont
  {D.~G.}\ \bibnamefont {Mandrus}}, \bibinfo {author} {\bibfnamefont
  {J.}~\bibnamefont {Yan}}, \bibinfo {author} {\bibfnamefont {W.}~\bibnamefont
  {Yao}}, \ and\ \bibinfo {author} {\bibfnamefont {X.}~\bibnamefont {Xu}},\
  }\href {\doibase 10.1038/s41586-019-0957-1} {\bibfield  {journal} {\bibinfo
  {journal} {Nature}\ }\textbf {\bibinfo {volume} {567}},\ \bibinfo {pages}
  {66} (\bibinfo {year} {2019})}\BibitemShut {NoStop}%
\bibitem [{\citenamefont {Jin}\ \emph {et~al.}(2019)\citenamefont {Jin},
  \citenamefont {Regan}, \citenamefont {Yan}, \citenamefont {Iqbal
  Bakti~Utama}, \citenamefont {Wang}, \citenamefont {Zhao}, \citenamefont
  {Qin}, \citenamefont {Yang}, \citenamefont {Zheng}, \citenamefont {Shi},
  \citenamefont {Watanabe}, \citenamefont {Taniguchi}, \citenamefont {Tongay},
  \citenamefont {Zettl},\ and\ \citenamefont {Wang}}]{Jin19}%
  \BibitemOpen
  \bibfield  {author} {\bibinfo {author} {\bibfnamefont {C.}~\bibnamefont
  {Jin}}, \bibinfo {author} {\bibfnamefont {E.~C.}\ \bibnamefont {Regan}},
  \bibinfo {author} {\bibfnamefont {A.}~\bibnamefont {Yan}}, \bibinfo {author}
  {\bibfnamefont {M.}~\bibnamefont {Iqbal Bakti~Utama}}, \bibinfo {author}
  {\bibfnamefont {D.}~\bibnamefont {Wang}}, \bibinfo {author} {\bibfnamefont
  {S.}~\bibnamefont {Zhao}}, \bibinfo {author} {\bibfnamefont {Y.}~\bibnamefont
  {Qin}}, \bibinfo {author} {\bibfnamefont {S.}~\bibnamefont {Yang}}, \bibinfo
  {author} {\bibfnamefont {Z.}~\bibnamefont {Zheng}}, \bibinfo {author}
  {\bibfnamefont {S.}~\bibnamefont {Shi}}, \bibinfo {author} {\bibfnamefont
  {K.}~\bibnamefont {Watanabe}}, \bibinfo {author} {\bibfnamefont
  {T.}~\bibnamefont {Taniguchi}}, \bibinfo {author} {\bibfnamefont
  {S.}~\bibnamefont {Tongay}}, \bibinfo {author} {\bibfnamefont
  {A.}~\bibnamefont {Zettl}}, \ and\ \bibinfo {author} {\bibfnamefont
  {F.}~\bibnamefont {Wang}},\ }\href {\doibase 10.1038/s41586-019-0976-y}
  {\bibfield  {journal} {\bibinfo  {journal} {Nature}\ }\textbf {\bibinfo
  {volume} {567}},\ \bibinfo {pages} {76} (\bibinfo {year} {2019})}\BibitemShut
  {NoStop}%
\bibitem [{\citenamefont {Alexeev}\ \emph {et~al.}(2019)\citenamefont
  {Alexeev}, \citenamefont {Ruiz-Tijerina}, \citenamefont {Danovich},
  \citenamefont {Hamer}, \citenamefont {Terry}, \citenamefont {Nayak},
  \citenamefont {Ahn}, \citenamefont {Pak}, \citenamefont {Lee}, \citenamefont
  {Sohn}, \citenamefont {Molas}, \citenamefont {Koperski}, \citenamefont
  {Watanabe}, \citenamefont {Taniguchi}, \citenamefont {Novoselov},
  \citenamefont {Gorbachev}, \citenamefont {Shin}, \citenamefont {Fal'ko},\
  and\ \citenamefont {Tartakovskii}}]{Ale19}%
  \BibitemOpen
  \bibfield  {author} {\bibinfo {author} {\bibfnamefont {E.~M.}\ \bibnamefont
  {Alexeev}}, \bibinfo {author} {\bibfnamefont {D.~A.}\ \bibnamefont
  {Ruiz-Tijerina}}, \bibinfo {author} {\bibfnamefont {M.}~\bibnamefont
  {Danovich}}, \bibinfo {author} {\bibfnamefont {M.~J.}\ \bibnamefont {Hamer}},
  \bibinfo {author} {\bibfnamefont {D.~J.}\ \bibnamefont {Terry}}, \bibinfo
  {author} {\bibfnamefont {P.~K.}\ \bibnamefont {Nayak}}, \bibinfo {author}
  {\bibfnamefont {S.}~\bibnamefont {Ahn}}, \bibinfo {author} {\bibfnamefont
  {S.}~\bibnamefont {Pak}}, \bibinfo {author} {\bibfnamefont {J.}~\bibnamefont
  {Lee}}, \bibinfo {author} {\bibfnamefont {J.~I.}\ \bibnamefont {Sohn}},
  \bibinfo {author} {\bibfnamefont {M.~R.}\ \bibnamefont {Molas}}, \bibinfo
  {author} {\bibfnamefont {M.}~\bibnamefont {Koperski}}, \bibinfo {author}
  {\bibfnamefont {K.}~\bibnamefont {Watanabe}}, \bibinfo {author}
  {\bibfnamefont {T.}~\bibnamefont {Taniguchi}}, \bibinfo {author}
  {\bibfnamefont {K.~S.}\ \bibnamefont {Novoselov}}, \bibinfo {author}
  {\bibfnamefont {R.~V.}\ \bibnamefont {Gorbachev}}, \bibinfo {author}
  {\bibfnamefont {H.~S.}\ \bibnamefont {Shin}}, \bibinfo {author}
  {\bibfnamefont {V.~I.}\ \bibnamefont {Fal'ko}}, \ and\ \bibinfo {author}
  {\bibfnamefont {A.~I.}\ \bibnamefont {Tartakovskii}},\ }\href {\doibase
  10.1038/s41586-019-0986-9} {\bibfield  {journal} {\bibinfo  {journal}
  {Nature}\ }\textbf {\bibinfo {volume} {567}},\ \bibinfo {pages} {81}
  (\bibinfo {year} {2019})}\BibitemShut {NoStop}%
\bibitem [{\citenamefont {Mermin}\ and\ \citenamefont {Wagner}(1966)}]{Mer66}%
  \BibitemOpen
  \bibfield  {author} {\bibinfo {author} {\bibfnamefont {N.~D.}\ \bibnamefont
  {Mermin}}\ and\ \bibinfo {author} {\bibfnamefont {H.}~\bibnamefont
  {Wagner}},\ }\href {\doibase 10.1103/PhysRevLett.17.1133} {\bibfield
  {journal} {\bibinfo  {journal} {Phys. Rev. Lett.}\ }\textbf {\bibinfo
  {volume} {17}},\ \bibinfo {pages} {1133} (\bibinfo {year}
  {1966})}\BibitemShut {NoStop}%
\bibitem [{\citenamefont {Huang}\ \emph {et~al.}(2017)\citenamefont {Huang},
  \citenamefont {Clark}, \citenamefont {Navarro-Moratalla}, \citenamefont
  {Klein}, \citenamefont {Cheng}, \citenamefont {Seyler}, \citenamefont
  {Zhong}, \citenamefont {Schmidgall}, \citenamefont {McGuire}, \citenamefont
  {Cobden}, \citenamefont {Yao}, \citenamefont {Xiao}, \citenamefont
  {Jarillo-Herrero},\ and\ \citenamefont {Xu}}]{huang_layer-dependent_2017}%
  \BibitemOpen
  \bibfield  {author} {\bibinfo {author} {\bibfnamefont {B.}~\bibnamefont
  {Huang}}, \bibinfo {author} {\bibfnamefont {G.}~\bibnamefont {Clark}},
  \bibinfo {author} {\bibfnamefont {E.}~\bibnamefont {Navarro-Moratalla}},
  \bibinfo {author} {\bibfnamefont {D.~R.}\ \bibnamefont {Klein}}, \bibinfo
  {author} {\bibfnamefont {R.}~\bibnamefont {Cheng}}, \bibinfo {author}
  {\bibfnamefont {K.~L.}\ \bibnamefont {Seyler}}, \bibinfo {author}
  {\bibfnamefont {D.}~\bibnamefont {Zhong}}, \bibinfo {author} {\bibfnamefont
  {E.}~\bibnamefont {Schmidgall}}, \bibinfo {author} {\bibfnamefont {M.~A.}\
  \bibnamefont {McGuire}}, \bibinfo {author} {\bibfnamefont {D.~H.}\
  \bibnamefont {Cobden}}, \bibinfo {author} {\bibfnamefont {W.}~\bibnamefont
  {Yao}}, \bibinfo {author} {\bibfnamefont {D.}~\bibnamefont {Xiao}}, \bibinfo
  {author} {\bibfnamefont {P.}~\bibnamefont {Jarillo-Herrero}}, \ and\ \bibinfo
  {author} {\bibfnamefont {X.}~\bibnamefont {Xu}},\ }\href {\doibase
  10.1038/nature22391} {\bibfield  {journal} {\bibinfo  {journal} {Nature}\
  }\textbf {\bibinfo {volume} {546}},\ \bibinfo {pages} {270} (\bibinfo {year}
  {2017})}\BibitemShut {NoStop}%
\bibitem [{\citenamefont {Burch}\ \emph {et~al.}(2018)\citenamefont {Burch},
  \citenamefont {Mandrus},\ and\ \citenamefont {Park}}]{Bur18}%
  \BibitemOpen
  \bibfield  {author} {\bibinfo {author} {\bibfnamefont {K.~S.}\ \bibnamefont
  {Burch}}, \bibinfo {author} {\bibfnamefont {D.}~\bibnamefont {Mandrus}}, \
  and\ \bibinfo {author} {\bibfnamefont {J.-G.}\ \bibnamefont {Park}},\ }\href
  {\doibase 10.1038/s41586-018-0631-z} {\bibfield  {journal} {\bibinfo
  {journal} {Nature}\ }\textbf {\bibinfo {volume} {563}},\ \bibinfo {pages}
  {47} (\bibinfo {year} {2018})}\BibitemShut {NoStop}%
\bibitem [{\citenamefont {Gibertini}\ \emph {et~al.}(2019)\citenamefont
  {Gibertini}, \citenamefont {Koperski}, \citenamefont {Morpurgo},\ and\
  \citenamefont {Novoselov}}]{Gib19}%
  \BibitemOpen
  \bibfield  {author} {\bibinfo {author} {\bibfnamefont {M.}~\bibnamefont
  {Gibertini}}, \bibinfo {author} {\bibfnamefont {M.}~\bibnamefont {Koperski}},
  \bibinfo {author} {\bibfnamefont {A.~F.}\ \bibnamefont {Morpurgo}}, \ and\
  \bibinfo {author} {\bibfnamefont {K.~S.}\ \bibnamefont {Novoselov}},\ }\href
  {\doibase 10.1038/s41565-019-0438-6} {\bibfield  {journal} {\bibinfo
  {journal} {Nature Nanotechnology}\ }\textbf {\bibinfo {volume} {14}},\
  \bibinfo {pages} {408} (\bibinfo {year} {2019})}\BibitemShut {NoStop}%
\bibitem [{\citenamefont {Alloul}\ \emph {et~al.}(2009)\citenamefont {Alloul},
  \citenamefont {Bobroff}, \citenamefont {Gabay},\ and\ \citenamefont
  {Hirschfeld}}]{All09}%
  \BibitemOpen
  \bibfield  {author} {\bibinfo {author} {\bibfnamefont {H.}~\bibnamefont
  {Alloul}}, \bibinfo {author} {\bibfnamefont {J.}~\bibnamefont {Bobroff}},
  \bibinfo {author} {\bibfnamefont {M.}~\bibnamefont {Gabay}}, \ and\ \bibinfo
  {author} {\bibfnamefont {P.~J.}\ \bibnamefont {Hirschfeld}},\ }\href
  {\doibase 10.1103/RevModPhys.81.45} {\bibfield  {journal} {\bibinfo
  {journal} {Rev. Mod. Phys.}\ }\textbf {\bibinfo {volume} {81}},\ \bibinfo
  {pages} {45} (\bibinfo {year} {2009})}\BibitemShut {NoStop}%
\bibitem [{\citenamefont {Praetorius}\ and\ \citenamefont
  {Fauth}(2017)}]{Pra17}%
  \BibitemOpen
  \bibfield  {author} {\bibinfo {author} {\bibfnamefont {C.}~\bibnamefont
  {Praetorius}}\ and\ \bibinfo {author} {\bibfnamefont {K.}~\bibnamefont
  {Fauth}},\ }\href {\doibase 10.1103/PhysRevB.95.115113} {\bibfield  {journal}
  {\bibinfo  {journal} {Phys. Rev. B}\ }\textbf {\bibinfo {volume} {95}},\
  \bibinfo {pages} {115113} (\bibinfo {year} {2017})}\BibitemShut {NoStop}%
\bibitem [{\citenamefont {{\v{C}}ervenka}\ \emph {et~al.}(2009)\citenamefont
  {{\v{C}}ervenka}, \citenamefont {Katsnelson},\ and\ \citenamefont
  {Flipse}}]{Cer09}%
  \BibitemOpen
  \bibfield  {author} {\bibinfo {author} {\bibfnamefont {J.}~\bibnamefont
  {{\v{C}}ervenka}}, \bibinfo {author} {\bibfnamefont {M.}~\bibnamefont
  {Katsnelson}}, \ and\ \bibinfo {author} {\bibfnamefont {C.}~\bibnamefont
  {Flipse}},\ }\href@noop {} {\bibfield  {journal} {\bibinfo  {journal} {Nature
  Physics}\ }\textbf {\bibinfo {volume} {5}},\ \bibinfo {pages} {840} (\bibinfo
  {year} {2009})}\BibitemShut {NoStop}%
\bibitem [{\citenamefont {Yazyev}\ and\ \citenamefont {Helm}(2007)}]{Yaz07}%
  \BibitemOpen
  \bibfield  {author} {\bibinfo {author} {\bibfnamefont {O.~V.}\ \bibnamefont
  {Yazyev}}\ and\ \bibinfo {author} {\bibfnamefont {L.}~\bibnamefont {Helm}},\
  }\href {\doibase 10.1103/PhysRevB.75.125408} {\bibfield  {journal} {\bibinfo
  {journal} {Phys. Rev. B}\ }\textbf {\bibinfo {volume} {75}},\ \bibinfo
  {pages} {125408} (\bibinfo {year} {2007})}\BibitemShut {NoStop}%
\bibitem [{\citenamefont {Bonilla}\ \emph {et~al.}(2018)\citenamefont
  {Bonilla}, \citenamefont {Kolekar}, \citenamefont {Ma}, \citenamefont {Diaz},
  \citenamefont {Kalappattil}, \citenamefont {Das}, \citenamefont {Eggers},
  \citenamefont {Gutierrez}, \citenamefont {Phan},\ and\ \citenamefont
  {Batzill}}]{bonilla_strong_2018}%
  \BibitemOpen
  \bibfield  {author} {\bibinfo {author} {\bibfnamefont {M.}~\bibnamefont
  {Bonilla}}, \bibinfo {author} {\bibfnamefont {S.}~\bibnamefont {Kolekar}},
  \bibinfo {author} {\bibfnamefont {Y.}~\bibnamefont {Ma}}, \bibinfo {author}
  {\bibfnamefont {H.~C.}\ \bibnamefont {Diaz}}, \bibinfo {author}
  {\bibfnamefont {V.}~\bibnamefont {Kalappattil}}, \bibinfo {author}
  {\bibfnamefont {R.}~\bibnamefont {Das}}, \bibinfo {author} {\bibfnamefont
  {T.}~\bibnamefont {Eggers}}, \bibinfo {author} {\bibfnamefont {H.~R.}\
  \bibnamefont {Gutierrez}}, \bibinfo {author} {\bibfnamefont {M.-H.}\
  \bibnamefont {Phan}}, \ and\ \bibinfo {author} {\bibfnamefont
  {M.}~\bibnamefont {Batzill}},\ }\href {\doibase 10.1038/s41565-018-0063-9}
  {\bibfield  {journal} {\bibinfo  {journal} {Nature Nanotechnology}\ }
  (\bibinfo {year} {2018}),\ 10.1038/s41565-018-0063-9}\BibitemShut {NoStop}%
\bibitem [{\citenamefont {Gong}\ \emph {et~al.}(2017)\citenamefont {Gong},
  \citenamefont {Li}, \citenamefont {Li}, \citenamefont {Ji}, \citenamefont
  {Stern}, \citenamefont {Xia}, \citenamefont {Cao}, \citenamefont {Bao},
  \citenamefont {Wang}, \citenamefont {Wang}, \citenamefont {Qiu},
  \citenamefont {Cava}, \citenamefont {Louie}, \citenamefont {Xia},\ and\
  \citenamefont {Zhang}}]{gong_discovery_2017}%
  \BibitemOpen
  \bibfield  {author} {\bibinfo {author} {\bibfnamefont {C.}~\bibnamefont
  {Gong}}, \bibinfo {author} {\bibfnamefont {L.}~\bibnamefont {Li}}, \bibinfo
  {author} {\bibfnamefont {Z.}~\bibnamefont {Li}}, \bibinfo {author}
  {\bibfnamefont {H.}~\bibnamefont {Ji}}, \bibinfo {author} {\bibfnamefont
  {A.}~\bibnamefont {Stern}}, \bibinfo {author} {\bibfnamefont
  {Y.}~\bibnamefont {Xia}}, \bibinfo {author} {\bibfnamefont {T.}~\bibnamefont
  {Cao}}, \bibinfo {author} {\bibfnamefont {W.}~\bibnamefont {Bao}}, \bibinfo
  {author} {\bibfnamefont {C.}~\bibnamefont {Wang}}, \bibinfo {author}
  {\bibfnamefont {Y.}~\bibnamefont {Wang}}, \bibinfo {author} {\bibfnamefont
  {Z.}~\bibnamefont {Qiu}}, \bibinfo {author} {\bibfnamefont {R.}~\bibnamefont
  {Cava}}, \bibinfo {author} {\bibfnamefont {S.~G.}\ \bibnamefont {Louie}},
  \bibinfo {author} {\bibfnamefont {J.}~\bibnamefont {Xia}}, \ and\ \bibinfo
  {author} {\bibfnamefont {X.}~\bibnamefont {Zhang}},\ }\href {\doibase
  10.1364/CLEO_AT.2017.JTh5C.2} {\ ,\ \bibinfo {pages} {JTh5C.2} (\bibinfo
  {year} {2017})}\BibitemShut {NoStop}%
\bibitem [{\citenamefont {Guo}\ \emph {et~al.}()\citenamefont {Guo},
  \citenamefont {Deng}, \citenamefont {Sun}, \citenamefont {Li}, \citenamefont
  {Zhao}, \citenamefont {Wu}, \citenamefont {Chu}, \citenamefont {Zhang},
  \citenamefont {Pan}, \citenamefont {Zheng}, \citenamefont {Wu}, \citenamefont
  {Jin}, \citenamefont {Wu},\ and\ \citenamefont {Xie}}]{VS2_FM}%
  \BibitemOpen
  \bibfield  {author} {\bibinfo {author} {\bibfnamefont {Y.}~\bibnamefont
  {Guo}}, \bibinfo {author} {\bibfnamefont {H.}~\bibnamefont {Deng}}, \bibinfo
  {author} {\bibfnamefont {X.}~\bibnamefont {Sun}}, \bibinfo {author}
  {\bibfnamefont {X.}~\bibnamefont {Li}}, \bibinfo {author} {\bibfnamefont
  {J.}~\bibnamefont {Zhao}}, \bibinfo {author} {\bibfnamefont {J.}~\bibnamefont
  {Wu}}, \bibinfo {author} {\bibfnamefont {W.}~\bibnamefont {Chu}}, \bibinfo
  {author} {\bibfnamefont {S.}~\bibnamefont {Zhang}}, \bibinfo {author}
  {\bibfnamefont {H.}~\bibnamefont {Pan}}, \bibinfo {author} {\bibfnamefont
  {X.}~\bibnamefont {Zheng}}, \bibinfo {author} {\bibfnamefont
  {X.}~\bibnamefont {Wu}}, \bibinfo {author} {\bibfnamefont {C.}~\bibnamefont
  {Jin}}, \bibinfo {author} {\bibfnamefont {C.}~\bibnamefont {Wu}}, \ and\
  \bibinfo {author} {\bibfnamefont {Y.}~\bibnamefont {Xie}},\ }\href {\doibase
  10.1002/adma.201700715} {\bibfield  {journal} {\bibinfo  {journal} {Advanced
  Materials}\ }\textbf {\bibinfo {volume} {29}},\ \bibinfo {pages}
  {1700715}}\BibitemShut {NoStop}%
\bibitem [{\citenamefont {Bussmann-Holder}\ and\ \citenamefont
  {Büttner}(2002)}]{TiSe2_CDW}%
  \BibitemOpen
  \bibfield  {author} {\bibinfo {author} {\bibfnamefont {A.}~\bibnamefont
  {Bussmann-Holder}}\ and\ \bibinfo {author} {\bibfnamefont {H.}~\bibnamefont
  {Büttner}},\ }\href {http://stacks.iop.org/0953-8984/14/i=34/a=316}
  {\bibfield  {journal} {\bibinfo  {journal} {Journal of Physics: Condensed
  Matter}\ }\textbf {\bibinfo {volume} {14}},\ \bibinfo {pages} {7973}
  (\bibinfo {year} {2002})}\BibitemShut {NoStop}%
\bibitem [{\citenamefont {Wilson}\ \emph {et~al.}(1975)\citenamefont {Wilson},
  \citenamefont {Salvo},\ and\ \citenamefont {Mahajan}}]{TMD_CDW}%
  \BibitemOpen
  \bibfield  {author} {\bibinfo {author} {\bibfnamefont {J.}~\bibnamefont
  {Wilson}}, \bibinfo {author} {\bibfnamefont {F.~D.}\ \bibnamefont {Salvo}}, \
  and\ \bibinfo {author} {\bibfnamefont {S.}~\bibnamefont {Mahajan}},\ }\href
  {\doibase 10.1080/00018737500101391} {\bibfield  {journal} {\bibinfo
  {journal} {Advances in Physics}\ }\textbf {\bibinfo {volume} {24}},\ \bibinfo
  {pages} {117} (\bibinfo {year} {1975})},\ \Eprint
  {http://arxiv.org/abs/https://doi.org/10.1080/00018737500101391}
  {https://doi.org/10.1080/00018737500101391} \BibitemShut {NoStop}%
\bibitem [{\citenamefont {Van~Bruggen}\ and\ \citenamefont
  {Haas}(1976)}]{van_bruggen_magnetic_1976}%
  \BibitemOpen
  \bibfield  {author} {\bibinfo {author} {\bibfnamefont {C.~F.}\ \bibnamefont
  {Van~Bruggen}}\ and\ \bibinfo {author} {\bibfnamefont {C.}~\bibnamefont
  {Haas}},\ }\href@noop {} {\bibfield  {journal} {\bibinfo  {journal} {Solid
  State Communications}\ }\textbf {\bibinfo {volume} {20}},\ \bibinfo {pages}
  {251} (\bibinfo {year} {1976})}\BibitemShut {NoStop}%
\bibitem [{\citenamefont {Yadav}\ and\ \citenamefont
  {Rastogi}(2010)}]{yadav_electronic_2010}%
  \BibitemOpen
  \bibfield  {author} {\bibinfo {author} {\bibfnamefont {C.}~\bibnamefont
  {Yadav}}\ and\ \bibinfo {author} {\bibfnamefont {A.}~\bibnamefont
  {Rastogi}},\ }\href {\doibase 10.1016/j.ssc.2009.12.026} {\bibfield
  {journal} {\bibinfo  {journal} {Solid State Communications}\ }\textbf
  {\bibinfo {volume} {150}},\ \bibinfo {pages} {648} (\bibinfo {year}
  {2010})}\BibitemShut {NoStop}%
\bibitem [{\citenamefont {Williams}\ \emph {et~al.}(1976)\citenamefont
  {Williams}, \citenamefont {Scruby}, \citenamefont {Clark},\ and\
  \citenamefont {Parry}}]{williams_charge_1976}%
  \BibitemOpen
  \bibfield  {author} {\bibinfo {author} {\bibfnamefont {P.~M.}\ \bibnamefont
  {Williams}}, \bibinfo {author} {\bibfnamefont {C.~B.}\ \bibnamefont
  {Scruby}}, \bibinfo {author} {\bibfnamefont {W.~B.}\ \bibnamefont {Clark}}, \
  and\ \bibinfo {author} {\bibfnamefont {G.~S.}\ \bibnamefont {Parry}},\
  }\href@noop {} {\bibfield  {journal} {\bibinfo  {journal} {Le Journal de
  Physique Colloques}\ }\textbf {\bibinfo {volume} {37}},\ \bibinfo {pages}
  {C4} (\bibinfo {year} {1976})}\BibitemShut {NoStop}%
\bibitem [{\citenamefont {Eaglesham}\ \emph {et~al.}(1986)\citenamefont
  {Eaglesham}, \citenamefont {Withers},\ and\ \citenamefont
  {Bird}}]{VSe2_CDW1}%
  \BibitemOpen
  \bibfield  {author} {\bibinfo {author} {\bibfnamefont {D.~J.}\ \bibnamefont
  {Eaglesham}}, \bibinfo {author} {\bibfnamefont {R.~L.}\ \bibnamefont
  {Withers}}, \ and\ \bibinfo {author} {\bibfnamefont {D.~M.}\ \bibnamefont
  {Bird}},\ }\href {http://stacks.iop.org/0022-3719/19/i=3/a=006} {\bibfield
  {journal} {\bibinfo  {journal} {Journal of Physics C: Solid State Physics}\
  }\textbf {\bibinfo {volume} {19}},\ \bibinfo {pages} {359} (\bibinfo {year}
  {1986})}\BibitemShut {NoStop}%
\bibitem [{\citenamefont {Woolley}\ and\ \citenamefont
  {Wexler}(1977)}]{0022-3719-10-14-013}%
  \BibitemOpen
  \bibfield  {author} {\bibinfo {author} {\bibfnamefont {A.~M.}\ \bibnamefont
  {Woolley}}\ and\ \bibinfo {author} {\bibfnamefont {G.}~\bibnamefont
  {Wexler}},\ }\href {http://stacks.iop.org/0022-3719/10/i=14/a=013} {\bibfield
   {journal} {\bibinfo  {journal} {Journal of Physics C: Solid State Physics}\
  }\textbf {\bibinfo {volume} {10}},\ \bibinfo {pages} {2601} (\bibinfo {year}
  {1977})}\BibitemShut {NoStop}%
\bibitem [{\citenamefont {Strocov}\ \emph {et~al.}(2012)\citenamefont
  {Strocov}, \citenamefont {Shi}, \citenamefont {Kobayashi}, \citenamefont
  {Monney}, \citenamefont {Wang}, \citenamefont {Krempasky}, \citenamefont
  {Schmitt}, \citenamefont {Patthey}, \citenamefont {Berger},\ and\
  \citenamefont {Blaha}}]{PhysRevLett.109.086401}%
  \BibitemOpen
  \bibfield  {author} {\bibinfo {author} {\bibfnamefont {V.~N.}\ \bibnamefont
  {Strocov}}, \bibinfo {author} {\bibfnamefont {M.}~\bibnamefont {Shi}},
  \bibinfo {author} {\bibfnamefont {M.}~\bibnamefont {Kobayashi}}, \bibinfo
  {author} {\bibfnamefont {C.}~\bibnamefont {Monney}}, \bibinfo {author}
  {\bibfnamefont {X.}~\bibnamefont {Wang}}, \bibinfo {author} {\bibfnamefont
  {J.}~\bibnamefont {Krempasky}}, \bibinfo {author} {\bibfnamefont
  {T.}~\bibnamefont {Schmitt}}, \bibinfo {author} {\bibfnamefont
  {L.}~\bibnamefont {Patthey}}, \bibinfo {author} {\bibfnamefont
  {H.}~\bibnamefont {Berger}}, \ and\ \bibinfo {author} {\bibfnamefont
  {P.}~\bibnamefont {Blaha}},\ }\href {\doibase 10.1103/PhysRevLett.109.086401}
  {\bibfield  {journal} {\bibinfo  {journal} {Phys. Rev. Lett.}\ }\textbf
  {\bibinfo {volume} {109}},\ \bibinfo {pages} {086401} (\bibinfo {year}
  {2012})}\BibitemShut {NoStop}%
\bibitem [{\citenamefont {Bayard}\ and\ \citenamefont
  {Sienko}(1976)}]{BAYARD1976325}%
  \BibitemOpen
  \bibfield  {author} {\bibinfo {author} {\bibfnamefont {M.}~\bibnamefont
  {Bayard}}\ and\ \bibinfo {author} {\bibfnamefont {M.}~\bibnamefont
  {Sienko}},\ }\href {\doibase https://doi.org/10.1016/0022-4596(76)90184-5}
  {\bibfield  {journal} {\bibinfo  {journal} {Journal of Solid State
  Chemistry}\ }\textbf {\bibinfo {volume} {19}},\ \bibinfo {pages} {325 }
  (\bibinfo {year} {1976})}\BibitemShut {NoStop}%
\bibitem [{\citenamefont {Barua}\ \emph
  {et~al.}(2017{\natexlab{a}})\citenamefont {Barua}, \citenamefont {Hatnean},
  \citenamefont {Lees},\ and\ \citenamefont
  {Balakrishnan}}]{barua_signatures_2017}%
  \BibitemOpen
  \bibfield  {author} {\bibinfo {author} {\bibfnamefont {S.}~\bibnamefont
  {Barua}}, \bibinfo {author} {\bibfnamefont {M.~C.}\ \bibnamefont {Hatnean}},
  \bibinfo {author} {\bibfnamefont {M.~R.}\ \bibnamefont {Lees}}, \ and\
  \bibinfo {author} {\bibfnamefont {G.}~\bibnamefont {Balakrishnan}},\ }\href
  {\doibase 10.1038/s41598-017-11247-4} {\bibfield  {journal} {\bibinfo
  {journal} {Scientific Reports}\ }\textbf {\bibinfo {volume} {7}} (\bibinfo
  {year} {2017}{\natexlab{a}}),\ 10.1038/s41598-017-11247-4}\BibitemShut
  {NoStop}%
\bibitem [{\citenamefont {Cao}\ \emph {et~al.}(2017)\citenamefont {Cao},
  \citenamefont {Yun}, \citenamefont {Sang}, \citenamefont {Xiang},
  \citenamefont {Liu},\ and\ \citenamefont {Wang}}]{cao_defect_2017}%
  \BibitemOpen
  \bibfield  {author} {\bibinfo {author} {\bibfnamefont {Q.}~\bibnamefont
  {Cao}}, \bibinfo {author} {\bibfnamefont {F.~F.}\ \bibnamefont {Yun}},
  \bibinfo {author} {\bibfnamefont {L.}~\bibnamefont {Sang}}, \bibinfo {author}
  {\bibfnamefont {F.}~\bibnamefont {Xiang}}, \bibinfo {author} {\bibfnamefont
  {G.}~\bibnamefont {Liu}}, \ and\ \bibinfo {author} {\bibfnamefont
  {X.}~\bibnamefont {Wang}},\ }\href {\doibase 10.1088/1361-6528/aa8f6c}
  {\bibfield  {journal} {\bibinfo  {journal} {Nanotechnology}\ }\textbf
  {\bibinfo {volume} {28}},\ \bibinfo {pages} {475703} (\bibinfo {year}
  {2017})}\BibitemShut {NoStop}%
\bibitem [{\citenamefont {Li}\ \emph {et~al.}(2014)\citenamefont {Li},
  \citenamefont {Tu},\ and\ \citenamefont {Chen}}]{li_versatile_2014}%
  \BibitemOpen
  \bibfield  {author} {\bibinfo {author} {\bibfnamefont {F.}~\bibnamefont
  {Li}}, \bibinfo {author} {\bibfnamefont {K.}~\bibnamefont {Tu}}, \ and\
  \bibinfo {author} {\bibfnamefont {Z.}~\bibnamefont {Chen}},\ }\href {\doibase
  10.1021/jp507093t} {\bibfield  {journal} {\bibinfo  {journal} {The Journal of
  Physical Chemistry C}\ }\textbf {\bibinfo {volume} {118}},\ \bibinfo {pages}
  {21264} (\bibinfo {year} {2014})}\BibitemShut {NoStop}%
\bibitem [{\citenamefont {Ma}\ \emph {et~al.}(2012)\citenamefont {Ma},
  \citenamefont {Dai}, \citenamefont {Guo}, \citenamefont {Niu}, \citenamefont
  {Zhu},\ and\ \citenamefont {Huang}}]{ma_evidence_2012}%
  \BibitemOpen
  \bibfield  {author} {\bibinfo {author} {\bibfnamefont {Y.}~\bibnamefont
  {Ma}}, \bibinfo {author} {\bibfnamefont {Y.}~\bibnamefont {Dai}}, \bibinfo
  {author} {\bibfnamefont {M.}~\bibnamefont {Guo}}, \bibinfo {author}
  {\bibfnamefont {C.}~\bibnamefont {Niu}}, \bibinfo {author} {\bibfnamefont
  {Y.}~\bibnamefont {Zhu}}, \ and\ \bibinfo {author} {\bibfnamefont
  {B.}~\bibnamefont {Huang}},\ }\href {\doibase 10.1021/nn204667z} {\bibfield
  {journal} {\bibinfo  {journal} {ACS Nano}\ }\textbf {\bibinfo {volume} {6}},\
  \bibinfo {pages} {1695} (\bibinfo {year} {2012})}\BibitemShut {NoStop}%
\bibitem [{\citenamefont {Esters}\ \emph {et~al.}(2017)\citenamefont {Esters},
  \citenamefont {Hennig},\ and\ \citenamefont {Johnson}}]{PhysRevB_DI}%
  \BibitemOpen
  \bibfield  {author} {\bibinfo {author} {\bibfnamefont {M.}~\bibnamefont
  {Esters}}, \bibinfo {author} {\bibfnamefont {R.~G.}\ \bibnamefont {Hennig}},
  \ and\ \bibinfo {author} {\bibfnamefont {D.~C.}\ \bibnamefont {Johnson}},\
  }\href {\doibase 10.1103/PhysRevB.96.235147} {\bibfield  {journal} {\bibinfo
  {journal} {Phys. Rev. B}\ }\textbf {\bibinfo {volume} {96}},\ \bibinfo
  {pages} {235147} (\bibinfo {year} {2017})}\BibitemShut {NoStop}%
\bibitem [{\citenamefont {Fuh}\ \emph {et~al.}(2016)\citenamefont {Fuh},
  \citenamefont {Yan}, \citenamefont {Wu}, \citenamefont {Felser},\ and\
  \citenamefont {Chang}}]{NJP}%
  \BibitemOpen
  \bibfield  {author} {\bibinfo {author} {\bibfnamefont {H.-R.}\ \bibnamefont
  {Fuh}}, \bibinfo {author} {\bibfnamefont {B.}~\bibnamefont {Yan}}, \bibinfo
  {author} {\bibfnamefont {S.-C.}\ \bibnamefont {Wu}}, \bibinfo {author}
  {\bibfnamefont {C.}~\bibnamefont {Felser}}, \ and\ \bibinfo {author}
  {\bibfnamefont {C.-R.}\ \bibnamefont {Chang}},\ }\href {\doibase
  10.1088/1367-2630/18/11/113038} {\bibfield  {journal} {\bibinfo  {journal}
  {New Journal of Physics}\ }\textbf {\bibinfo {volume} {18}},\ \bibinfo
  {pages} {113038} (\bibinfo {year} {2016})}\BibitemShut {NoStop}%
\bibitem [{\citenamefont {Barla}\ \emph {et~al.}(2016)\citenamefont {Barla},
  \citenamefont {Nicol{\'{a}}s}, \citenamefont {Cocco}, \citenamefont
  {Valvidares}, \citenamefont {Herrero-Mart{\'\i}n}, \citenamefont {Gargiani},
  \citenamefont {Moldes}, \citenamefont {Ruget}, \citenamefont {Pellegrin},\
  and\ \citenamefont {Ferrer}}]{Bar16}%
  \BibitemOpen
  \bibfield  {author} {\bibinfo {author} {\bibfnamefont {A.}~\bibnamefont
  {Barla}}, \bibinfo {author} {\bibfnamefont {J.}~\bibnamefont
  {Nicol{\'{a}}s}}, \bibinfo {author} {\bibfnamefont {D.}~\bibnamefont
  {Cocco}}, \bibinfo {author} {\bibfnamefont {S.~M.}\ \bibnamefont
  {Valvidares}}, \bibinfo {author} {\bibfnamefont {J.}~\bibnamefont
  {Herrero-Mart{\'\i}n}}, \bibinfo {author} {\bibfnamefont {P.}~\bibnamefont
  {Gargiani}}, \bibinfo {author} {\bibfnamefont {J.}~\bibnamefont {Moldes}},
  \bibinfo {author} {\bibfnamefont {C.}~\bibnamefont {Ruget}}, \bibinfo
  {author} {\bibfnamefont {E.}~\bibnamefont {Pellegrin}}, \ and\ \bibinfo
  {author} {\bibfnamefont {S.}~\bibnamefont {Ferrer}},\ }\href {\doibase
  10.1107/S1600577516013461} {\bibfield  {journal} {\bibinfo  {journal}
  {Journal of Synchrotron Radiation}\ }\textbf {\bibinfo {volume} {23}},\
  \bibinfo {pages} {1507} (\bibinfo {year} {2016})}\BibitemShut {NoStop}%
\bibitem [{\citenamefont {Haverkort}\ \emph {et~al.}(2012)\citenamefont
  {Haverkort}, \citenamefont {Zwierzycki},\ and\ \citenamefont
  {Andersen}}]{Hav12}%
  \BibitemOpen
  \bibfield  {author} {\bibinfo {author} {\bibfnamefont {M.~W.}\ \bibnamefont
  {Haverkort}}, \bibinfo {author} {\bibfnamefont {M.}~\bibnamefont
  {Zwierzycki}}, \ and\ \bibinfo {author} {\bibfnamefont {O.~K.}\ \bibnamefont
  {Andersen}},\ }\href {\doibase 10.1103/PhysRevB.85.165113} {\bibfield
  {journal} {\bibinfo  {journal} {Phys. Rev. B}\ }\textbf {\bibinfo {volume}
  {85}},\ \bibinfo {pages} {165113} (\bibinfo {year} {2012})}\BibitemShut
  {NoStop}%
\bibitem [{\citenamefont {Lu}\ \emph {et~al.}(2014)\citenamefont {Lu},
  \citenamefont {H\"oppner}, \citenamefont {Gunnarsson},\ and\ \citenamefont
  {Haverkort}}]{Lu14}%
  \BibitemOpen
  \bibfield  {author} {\bibinfo {author} {\bibfnamefont {Y.}~\bibnamefont
  {Lu}}, \bibinfo {author} {\bibfnamefont {M.}~\bibnamefont {H\"oppner}},
  \bibinfo {author} {\bibfnamefont {O.}~\bibnamefont {Gunnarsson}}, \ and\
  \bibinfo {author} {\bibfnamefont {M.~W.}\ \bibnamefont {Haverkort}},\ }\href
  {\doibase 10.1103/PhysRevB.90.085102} {\bibfield  {journal} {\bibinfo
  {journal} {Phys. Rev. B}\ }\textbf {\bibinfo {volume} {90}},\ \bibinfo
  {pages} {085102} (\bibinfo {year} {2014})}\BibitemShut {NoStop}%
\bibitem [{\citenamefont {Hohenberg}\ and\ \citenamefont {Kohn}(1964)}]{HK}%
  \BibitemOpen
  \bibfield  {author} {\bibinfo {author} {\bibfnamefont {P.}~\bibnamefont
  {Hohenberg}}\ and\ \bibinfo {author} {\bibfnamefont {W.}~\bibnamefont
  {Kohn}},\ }\href@noop {} {\bibfield  {journal} {\bibinfo  {journal} {Phys.
  Rev.}\ }\textbf {\bibinfo {volume} {136}},\ \bibinfo {pages} {B864} (\bibinfo
  {year} {1964})}\BibitemShut {NoStop}%
\bibitem [{\citenamefont {Kohn}\ and\ \citenamefont {Sham}(1965)}]{KS}%
  \BibitemOpen
  \bibfield  {author} {\bibinfo {author} {\bibfnamefont {W.}~\bibnamefont
  {Kohn}}\ and\ \bibinfo {author} {\bibfnamefont {L.~J.}\ \bibnamefont
  {Sham}},\ }\href@noop {} {\bibfield  {journal} {\bibinfo  {journal} {Phys.
  Rev.}\ }\textbf {\bibinfo {volume} {140}},\ \bibinfo {pages} {A1133}
  (\bibinfo {year} {1965})}\BibitemShut {NoStop}%
\bibitem [{\citenamefont {Schwarz}\ and\ \citenamefont {Blaha}(2003)}]{WIEN2k}%
  \BibitemOpen
  \bibfield  {author} {\bibinfo {author} {\bibfnamefont {K.}~\bibnamefont
  {Schwarz}}\ and\ \bibinfo {author} {\bibfnamefont {P.}~\bibnamefont
  {Blaha}},\ }\href@noop {} {\bibfield  {journal} {\bibinfo  {journal} {Comp.
  Mater. Sci.}\ }\textbf {\bibinfo {volume} {28}},\ \bibinfo {pages} {259}
  (\bibinfo {year} {2003})}\BibitemShut {NoStop}%
\bibitem [{\citenamefont {Perdew}\ \emph {et~al.}(1996)\citenamefont {Perdew},
  \citenamefont {Burke},\ and\ \citenamefont {Ernzerhof}}]{PBE}%
  \BibitemOpen
  \bibfield  {author} {\bibinfo {author} {\bibfnamefont {J.~P.}\ \bibnamefont
  {Perdew}}, \bibinfo {author} {\bibfnamefont {K.}~\bibnamefont {Burke}}, \
  and\ \bibinfo {author} {\bibfnamefont {M.}~\bibnamefont {Ernzerhof}},\
  }\href@noop {} {\bibfield  {journal} {\bibinfo  {journal} {Phys. Rev. Lett.}\
  }\textbf {\bibinfo {volume} {77}},\ \bibinfo {pages} {3865} (\bibinfo {year}
  {1996})}\BibitemShut {NoStop}%
\bibitem [{\citenamefont {Anisimov}\ \emph {et~al.}(1997)\citenamefont
  {Anisimov}, \citenamefont {Aryasetiawan},\ and\ \citenamefont
  {Lichtenstein}}]{LDAU}%
  \BibitemOpen
  \bibfield  {author} {\bibinfo {author} {\bibfnamefont {V.~I.}\ \bibnamefont
  {Anisimov}}, \bibinfo {author} {\bibfnamefont {F.}~\bibnamefont
  {Aryasetiawan}}, \ and\ \bibinfo {author} {\bibfnamefont {A.~I.}\
  \bibnamefont {Lichtenstein}},\ }\href@noop {} {\bibfield  {journal} {\bibinfo
   {journal} {J. Phys.: Condens. Mat.}\ }\textbf {\bibinfo {volume} {9}},\
  \bibinfo {pages} {767} (\bibinfo {year} {1997})}\BibitemShut {NoStop}%
\bibitem [{\citenamefont {Madsen}\ \emph {et~al.}(2018)\citenamefont {Madsen},
  \citenamefont {Carrete},\ and\ \citenamefont {Verstraete}}]{Boltztrap2}%
  \BibitemOpen
  \bibfield  {author} {\bibinfo {author} {\bibfnamefont {G.~K.}\ \bibnamefont
  {Madsen}}, \bibinfo {author} {\bibfnamefont {J.}~\bibnamefont {Carrete}}, \
  and\ \bibinfo {author} {\bibfnamefont {M.~J.}\ \bibnamefont {Verstraete}},\
  }\href {\doibase https://doi.org/10.1016/j.cpc.2018.05.010} {\bibfield
  {journal} {\bibinfo  {journal} {Computer Physics Communications}\ }\textbf
  {\bibinfo {volume} {231}},\ \bibinfo {pages} {140 } (\bibinfo {year}
  {2018})}\BibitemShut {NoStop}%
\bibitem [{\citenamefont {Barua}\ \emph
  {et~al.}(2017{\natexlab{b}})\citenamefont {Barua}, \citenamefont {Hatnean},
  \citenamefont {Lees},\ and\ \citenamefont {Balakrishnan}}]{Bar17}%
  \BibitemOpen
  \bibfield  {author} {\bibinfo {author} {\bibfnamefont {S.}~\bibnamefont
  {Barua}}, \bibinfo {author} {\bibfnamefont {M.~C.}\ \bibnamefont {Hatnean}},
  \bibinfo {author} {\bibfnamefont {M.~R.}\ \bibnamefont {Lees}}, \ and\
  \bibinfo {author} {\bibfnamefont {G.}~\bibnamefont {Balakrishnan}},\ }\href
  {\doibase 10.1038/s41598-017-11247-4} {\bibfield  {journal} {\bibinfo
  {journal} {Scientific Reports}\ }\textbf {\bibinfo {volume} {7}},\ \bibinfo
  {pages} {10964} (\bibinfo {year} {2017}{\natexlab{b}})}\BibitemShut {NoStop}%
\bibitem [{\citenamefont {Gr\"{u}ner}(1994)}]{Gru94}%
  \BibitemOpen
  \bibfield  {author} {\bibinfo {author} {\bibfnamefont {G.}~\bibnamefont
  {Gr\"{u}ner}},\ }\href@noop {} {\emph {\bibinfo {title} {Charge Density Waves
  in Solids}}}\ (\bibinfo  {publisher} {Addison-Wesley},\ \bibinfo {year}
  {1994})\BibitemShut {NoStop}%
\bibitem [{\citenamefont {Terashima}\ \emph {et~al.}(2003)\citenamefont
  {Terashima}, \citenamefont {Sato}, \citenamefont {Komatsu}, \citenamefont
  {Takahashi}, \citenamefont {Maeda},\ and\ \citenamefont
  {Hayashi}}]{terashima_charge-density_2003}%
  \BibitemOpen
  \bibfield  {author} {\bibinfo {author} {\bibfnamefont {K.}~\bibnamefont
  {Terashima}}, \bibinfo {author} {\bibfnamefont {T.}~\bibnamefont {Sato}},
  \bibinfo {author} {\bibfnamefont {H.}~\bibnamefont {Komatsu}}, \bibinfo
  {author} {\bibfnamefont {T.}~\bibnamefont {Takahashi}}, \bibinfo {author}
  {\bibfnamefont {N.}~\bibnamefont {Maeda}}, \ and\ \bibinfo {author}
  {\bibfnamefont {K.}~\bibnamefont {Hayashi}},\ }\href {\doibase
  10.1103/PhysRevB.68.155108} {\bibfield  {journal} {\bibinfo  {journal}
  {Physical Review B}\ }\textbf {\bibinfo {volume} {68}} (\bibinfo {year}
  {2003}),\ 10.1103/PhysRevB.68.155108}\BibitemShut {NoStop}%
\bibitem [{\citenamefont {Stoner}(1938)}]{Stoner_crit}%
  \BibitemOpen
  \bibfield  {author} {\bibinfo {author} {\bibfnamefont {E.~C.}\ \bibnamefont
  {Stoner}},\ }\href {\doibase 10.1098/rspa.1938.0066} {\bibfield  {journal}
  {\bibinfo  {journal} {Proceedings of the Royal Society of London A:
  Mathematical, Physical and Engineering Sciences}\ }\textbf {\bibinfo {volume}
  {165}},\ \bibinfo {pages} {372} (\bibinfo {year} {1938})}\BibitemShut
  {NoStop}%
\bibitem [{\citenamefont {Moriya}(1985)}]{moriya1985spin}%
  \BibitemOpen
  \bibfield  {author} {\bibinfo {author} {\bibfnamefont {T.}~\bibnamefont
  {Moriya}},\ }\href {https://books.google.es/books?id=PUssAAAAYAAJ} {\emph
  {\bibinfo {title} {Spin fluctuations in itinerant electron magnetism}}},\
  Springer Series in Solid-State Sciences\ (\bibinfo  {publisher}
  {Springer-Verlag},\ \bibinfo {year} {1985})\BibitemShut {NoStop}%
\bibitem [{\citenamefont {Fritsche}\ and\ \citenamefont
  {Weimert}()}]{stoner_atomicnumber}%
  \BibitemOpen
  \bibfield  {author} {\bibinfo {author} {\bibfnamefont {L.}~\bibnamefont
  {Fritsche}}\ and\ \bibinfo {author} {\bibfnamefont {B.}~\bibnamefont
  {Weimert}},\ }\href {\doibase
  10.1002/(SICI)1521-3951(199808)208:2<287::AID-PSSB287>3.0.CO;2-1} {\bibfield
  {journal} {\bibinfo  {journal} {physica status solidi (b)}\ }\textbf
  {\bibinfo {volume} {208}},\ \bibinfo {pages} {287}}\BibitemShut {NoStop}%
\bibitem [{\citenamefont {Zhang}\ \emph {et~al.}(2019)\citenamefont {Zhang},
  \citenamefont {Zhang}, \citenamefont {Wong}, \citenamefont {Yuan},
  \citenamefont {Vinai}, \citenamefont {Torelli}, \citenamefont {van~der Laan},
  \citenamefont {Feng},\ and\ \citenamefont {Wee}}]{Zhang2019}%
  \BibitemOpen
  \bibfield  {author} {\bibinfo {author} {\bibfnamefont {W.}~\bibnamefont
  {Zhang}}, \bibinfo {author} {\bibfnamefont {L.}~\bibnamefont {Zhang}},
  \bibinfo {author} {\bibfnamefont {P.~K.~J.}\ \bibnamefont {Wong}}, \bibinfo
  {author} {\bibfnamefont {J.}~\bibnamefont {Yuan}}, \bibinfo {author}
  {\bibfnamefont {G.}~\bibnamefont {Vinai}}, \bibinfo {author} {\bibfnamefont
  {P.}~\bibnamefont {Torelli}}, \bibinfo {author} {\bibfnamefont
  {G.}~\bibnamefont {van~der Laan}}, \bibinfo {author} {\bibfnamefont {Y.~P.}\
  \bibnamefont {Feng}}, \ and\ \bibinfo {author} {\bibfnamefont {A.~T.~S.}\
  \bibnamefont {Wee}},\ }\href {\doibase 10.1021/acsnano.9b02996} {\bibfield
  {journal} {\bibinfo  {journal} {ACS Nano}\ } (\bibinfo {year} {2019}),\
  10.1021/acsnano.9b02996}\BibitemShut {NoStop}%
\bibitem [{\citenamefont {Wong}\ \emph {et~al.}(2019)\citenamefont {Wong},
  \citenamefont {Zhang}, \citenamefont {Bussolotti}, \citenamefont {Yin},
  \citenamefont {Herng}, \citenamefont {Zhang}, \citenamefont {Huang},
  \citenamefont {Vinai}, \citenamefont {Krishnamurthi}, \citenamefont
  {Bukhvalov}, \citenamefont {Zheng}, \citenamefont {Chua}, \citenamefont
  {N'Diaye}, \citenamefont {Morton}, \citenamefont {Yang}, \citenamefont
  {Ou~Yang}, \citenamefont {Torelli}, \citenamefont {Chen}, \citenamefont
  {Goh}, \citenamefont {Ding}, \citenamefont {Lin}, \citenamefont {Brocks},
  \citenamefont {de~Jong}, \citenamefont {Castro~Neto},\ and\ \citenamefont
  {Wee}}]{Wong2019}%
  \BibitemOpen
  \bibfield  {author} {\bibinfo {author} {\bibfnamefont {P.~K.~J.}\
  \bibnamefont {Wong}}, \bibinfo {author} {\bibfnamefont {W.}~\bibnamefont
  {Zhang}}, \bibinfo {author} {\bibfnamefont {F.}~\bibnamefont {Bussolotti}},
  \bibinfo {author} {\bibfnamefont {X.}~\bibnamefont {Yin}}, \bibinfo {author}
  {\bibfnamefont {T.~S.}\ \bibnamefont {Herng}}, \bibinfo {author}
  {\bibfnamefont {L.}~\bibnamefont {Zhang}}, \bibinfo {author} {\bibfnamefont
  {Y.~L.}\ \bibnamefont {Huang}}, \bibinfo {author} {\bibfnamefont
  {G.}~\bibnamefont {Vinai}}, \bibinfo {author} {\bibfnamefont
  {S.}~\bibnamefont {Krishnamurthi}}, \bibinfo {author} {\bibfnamefont {D.~W.}\
  \bibnamefont {Bukhvalov}}, \bibinfo {author} {\bibfnamefont {Y.~J.}\
  \bibnamefont {Zheng}}, \bibinfo {author} {\bibfnamefont {R.}~\bibnamefont
  {Chua}}, \bibinfo {author} {\bibfnamefont {A.~T.}\ \bibnamefont {N'Diaye}},
  \bibinfo {author} {\bibfnamefont {S.~A.}\ \bibnamefont {Morton}}, \bibinfo
  {author} {\bibfnamefont {C.-Y.}\ \bibnamefont {Yang}}, \bibinfo {author}
  {\bibfnamefont {K.-H.}\ \bibnamefont {Ou~Yang}}, \bibinfo {author}
  {\bibfnamefont {P.}~\bibnamefont {Torelli}}, \bibinfo {author} {\bibfnamefont
  {W.}~\bibnamefont {Chen}}, \bibinfo {author} {\bibfnamefont {K.~E.~J.}\
  \bibnamefont {Goh}}, \bibinfo {author} {\bibfnamefont {J.}~\bibnamefont
  {Ding}}, \bibinfo {author} {\bibfnamefont {M.-T.}\ \bibnamefont {Lin}},
  \bibinfo {author} {\bibfnamefont {G.}~\bibnamefont {Brocks}}, \bibinfo
  {author} {\bibfnamefont {M.~P.}\ \bibnamefont {de~Jong}}, \bibinfo {author}
  {\bibfnamefont {A.~H.}\ \bibnamefont {Castro~Neto}}, \ and\ \bibinfo {author}
  {\bibfnamefont {A.~T.~S.}\ \bibnamefont {Wee}},\ }\href {\doibase
  10.1002/adma.201901185} {\bibfield  {journal} {\bibinfo  {journal} {Advanced
  Materials}\ }\textbf {\bibinfo {volume} {31}},\ \bibinfo {pages} {1901185}
  (\bibinfo {year} {2019})},\ \Eprint
  {http://arxiv.org/abs/https://onlinelibrary.wiley.com/doi/pdf/10.1002/adma.201901185}
  {https://onlinelibrary.wiley.com/doi/pdf/10.1002/adma.201901185} \BibitemShut
  {NoStop}%
\bibitem [{\citenamefont {Haverkort}(2005)}]{Hav05}%
  \BibitemOpen
  \bibfield  {author} {\bibinfo {author} {\bibfnamefont {M.}~\bibnamefont
  {Haverkort}},\ }\emph {\bibinfo {title} {Spin and orbital degrees of freedom
  in transition metal oxides and oxide thin films studied by soft x-ray
  absorption spectroscopy}},\ \href {https://kups.ub.uni-koeln.de/1455/} {Ph.D.
  thesis},\ \bibinfo  {school} {Universit{\"a}t zu K{\"o}ln} (\bibinfo {year}
  {2005})\BibitemShut {NoStop}%
\bibitem [{\citenamefont {Coelho}\ \emph {et~al.}(2019)\citenamefont {Coelho},
  \citenamefont {Nguyen~Cong}, \citenamefont {Bonilla}, \citenamefont
  {Kolekar}, \citenamefont {Phan}, \citenamefont {Avila}, \citenamefont
  {Asensio}, \citenamefont {Oleynik},\ and\ \citenamefont
  {Batzill}}]{Coelho2019}%
  \BibitemOpen
  \bibfield  {author} {\bibinfo {author} {\bibfnamefont {P.~M.}\ \bibnamefont
  {Coelho}}, \bibinfo {author} {\bibfnamefont {K.}~\bibnamefont {Nguyen~Cong}},
  \bibinfo {author} {\bibfnamefont {M.}~\bibnamefont {Bonilla}}, \bibinfo
  {author} {\bibfnamefont {S.}~\bibnamefont {Kolekar}}, \bibinfo {author}
  {\bibfnamefont {M.-H.}\ \bibnamefont {Phan}}, \bibinfo {author}
  {\bibfnamefont {J.}~\bibnamefont {Avila}}, \bibinfo {author} {\bibfnamefont
  {M.~C.}\ \bibnamefont {Asensio}}, \bibinfo {author} {\bibfnamefont {I.~I.}\
  \bibnamefont {Oleynik}}, \ and\ \bibinfo {author} {\bibfnamefont
  {M.}~\bibnamefont {Batzill}},\ }\href {\doibase 10.1021/acs.jpcc.9b04281}
  {\bibfield  {journal} {\bibinfo  {journal} {The Journal of Physical Chemistry
  C}\ }\textbf {\bibinfo {volume} {123}},\ \bibinfo {pages} {14089} (\bibinfo
  {year} {2019})}\BibitemShut {NoStop}%
\bibitem [{\citenamefont {Johannes}\ and\ \citenamefont
  {Mazin}(2008)}]{PhysRevB.77.165135}%
  \BibitemOpen
  \bibfield  {author} {\bibinfo {author} {\bibfnamefont {M.~D.}\ \bibnamefont
  {Johannes}}\ and\ \bibinfo {author} {\bibfnamefont {I.~I.}\ \bibnamefont
  {Mazin}},\ }\href {\doibase 10.1103/PhysRevB.77.165135} {\bibfield  {journal}
  {\bibinfo  {journal} {Phys. Rev. B}\ }\textbf {\bibinfo {volume} {77}},\
  \bibinfo {pages} {165135} (\bibinfo {year} {2008})}\BibitemShut {NoStop}%
\bibitem [{\citenamefont {Avsar}\ \emph {et~al.}(2019)\citenamefont {Avsar},
  \citenamefont {Ciarrocchi}, \citenamefont {Pizzochero}, \citenamefont
  {Unuchek}, \citenamefont {Yazyev},\ and\ \citenamefont {Kis}}]{Avsar2019}%
  \BibitemOpen
  \bibfield  {author} {\bibinfo {author} {\bibfnamefont {A.}~\bibnamefont
  {Avsar}}, \bibinfo {author} {\bibfnamefont {A.}~\bibnamefont {Ciarrocchi}},
  \bibinfo {author} {\bibfnamefont {M.}~\bibnamefont {Pizzochero}}, \bibinfo
  {author} {\bibfnamefont {D.}~\bibnamefont {Unuchek}}, \bibinfo {author}
  {\bibfnamefont {O.~V.}\ \bibnamefont {Yazyev}}, \ and\ \bibinfo {author}
  {\bibfnamefont {A.}~\bibnamefont {Kis}},\ }\href {\doibase
  10.1038/s41565-019-0467-1} {\bibfield  {journal} {\bibinfo  {journal} {Nature
  Nanotechnology}\ }\textbf {\bibinfo {volume} {14}},\ \bibinfo {pages} {674}
  (\bibinfo {year} {2019})}\BibitemShut {NoStop}%
\end{thebibliography}

%

\end{document}